\documentclass[review,hidelinks]{elsarticle}
\usepackage{graphicx}
\usepackage{tabularx}
\usepackage{tikz}
\usepackage{float}
\usepackage{caption}
\usepackage{subcaption}
\usepackage{lineno,hyperref}
\usepackage{mathtools}
\modulolinenumbers[5]

\journal{Journal of \LaTeX\ Templates}

\begin{document}

\begin{frontmatter}

\title{Effect of Grid Sensitivity on the Performance of Wall Adapting SGS Models for LES of Swirling and Separating-Reattaching Flows}


\author{Nitish Arya}

\author{Ashoke De\corref{mycorrespondingauthor}}
\cortext[mycorrespondingauthor]{Corresponding author}
\ead{ashoke@iitk.ac.in}

\address{Indian Institute of Technology Kanpur, Kanpur 208016}

\begin{abstract}
\quad The present study assesses the performance of the Wall Adapting SGS models along with the Dynamic Smagorinsky model for flows involving separation, reattachment and swirl. Due to the simple geometry and wide application in a variety of engineering systems, the Backward-Facing Step (BFS) geometry and Confined Swirling Flow (CSF) geometry are invoked in the present case. The calculation of the SGS stresses employs three models, namely, the Dynamic Smagorinsky model, the Wall Adapting Local Eddy viscosity (WALE) model and the Dynamic WALE model. For studying the effect of the grid sensitivity, the simulations are performed over two sets of grids with different resolutions based on the non-dimensional wall distance parameter( $y^+$ ).Grids corresponding to $y^+$=70 and $y^+$=20 are employed for the subsonic flow over the BFS while grids corresponding to  $y^+$=40 and $y^+$=20 are employed for supersonic flow over the BFS and for confined swirling flow geometry. The validation against the experimental results includes the mean flow fields and the turbulent stresses obtained for each case. The results reveal that for the fine grid ($y^+$=20), the near wall eddy viscosity profile for the WALE model is better than both the Dynamic WALE and the Dynamic Smagorinsky model. The difference between the predictions of the coarse and fine grids for Dynamic Smagorinsky and the WALE model is high whereas, the Dynamic WALE model is almost insensitive to the grid resolutions considered for the present case. The mean velocity and pressure values as well as the turbulent quantities predicted by the Dynamic WALE model are closest to the experimental values for all the cases.
\end{abstract}


\end{frontmatter}

\section{Introduction}

\quad Large Eddy Simulation (LES) is commonly used in the study of turbulent flows in which large scales of motion are resolved and the effect of small scales is modeled with the help of a Subgrid Scale (SGS) model which usually employs an eddy viscosity assumption to model the SGS stress. The Smagorinsky constant which appears in the original Smagorinsky model \cite{smagorinsky1963general} assumes a fixed value for the whole domain and for every time step and also it is determined apriori. However, for different flow configurations the value of the Smagorinsky constant should change. This fact was considered by Germano et al. \cite{germano1991dynamic} and they proposed a Dynamic model in which the Smagorinsky constant was calculated based on the information from the flow field.  Even with the dynamically calculated Smagorinsky constant, there are certain inherent limitations of the model. The choice of the velocity scale used in the model deters it from correctly predicting the eddy viscosity in the regions where the vorticity is much larger than the irrotational strain. The eddy viscosity in the near wall region is over-predicted due to large values of the velocity gradient which is a major drawback since all the turbulent fluctuations and consequently the eddy viscosity should vanish near the wall. Further, the eddy viscosity follows O(1) profile near the wall as opposed to O($y^3$) profile \cite{mizushina1970eddy}.

\quad The Wall Adapting Local Eddy Viscosity (WALE) model overcomes the limitations of the Smagorinsky model \cite{nicoud1999subgrid}. The WALE model uses a different velocity scale for the calculation of eddy viscosity which enables it to predict accurate values in the regions of high vorticity as well as high irrotational strain. The construction of the velocity scale provides a correct near wall behavior as well as forces the eddy viscosity to vanish at the wall. The only limitation of the WALE model is the fixed model constant which can also be calculated dynamically using the Germano-Lily procedure \cite{lilly1992proposed}. It has been observed, however, that the models which have correct near wall behavior are sensitive to the filtering operation. Therefore, the dynamic procedure for the WALE model involves the calculation of a Shear and Vortex Sensor (SVS) \cite{toda2010dynamic}, and the subsequent sections provide the details.

\quad Though the wall adapting models seem to overcome the limitations of Smagorinsky model, LES for a flow with high Reynolds number is computationally prohibitive if the wall is being resolved completely. One approach to reduce the computational cost is to use RANS-LES hybrid models where RANS is used in the near wall regions and LES is used in the core region \cite{kumar2017investigation}. Another approach is to use LES in conjunction with wall models. The wall models create a smooth profile for eddy viscosity from the wall upto the first grid point which lies in the logarithmic region \cite{soni2017characterization}. Since the use of a highly refined mesh near complex geometries poses many issues, not to mention the exorbitant amount of computational cost, it is only wise to use relatively coarser grids to study practical engineering flows. It would be interesting to study the wall adapting models with relatively coarser grids (not resolving the entire wall) and observe the mean quantities as well as the turbulent stresses. 

\quad However, the choice of the grid for LES is somewhat tricky. For RANS simulations a grid convergence test is employed and the chosen grid offers no significant improvement in the results even after refining the grid further. Such a method is not feasible for LES, firstly because of the computational restrictions. Secondly, as the LES grid is further refined, the contribution of the SGS model shifts towards smaller scales until the LES converges to DNS. The accuracy of the LES is inhibted by many factors such as the numerical errors and the modeling errors which usually interact with each other \cite{kravchenko1997effect,geurts2002framework,celik2005index}. It is difficult to separate numerical and modeling errors which makes grid independence test on LES rather complicated \cite{klein2005attempt}. However, there are specific quality parameters \cite{celik2005index, klein2005attempt} which show that the grids employed for LES are accompanied with minimum error. This enables us to test the prediction of various SGS models on relatively coarser grids associated with complex geometries while minimising the errors

\quad The flow through complex geometries involve exciting features such as swirl, separation and reattachment. The study of these feautures for simple geometries is the first step towards simulating flow over complex geometries. For example, a Backward-Facing Step geometry used in the construction of a scramjet engine, involves flow separation at the step corner and reattachment at some point downstream of the step. Another example is the Confined Swirling Flow geometry which involves sudden expansion of a swirling flow which mimics the flow through an industrial furnace or the combustion chamber of a jet engine. Both the flow configurations are simple yet the flow physics associated with them is intriguing as well as challenging, which makes them important for the validation of various numerical models.

\quad Thus, the present work assesses the performance of three SGS models viz.:  Dynamic Smagorinsky model, WALE model and Dynamic WALE model for three flow configurations - subsonic flow over a Backward Facing Step, supersonic flow over a Backward Facing Step and subsonic Confined Swirling Flow. Grids corresponding to $y^+$=70 and $y^+$=20 are employed for the subsonic flow over the BFS while grids corresponding to $y^+$=40 and $y^+$=20 are employed for the Confined Swirling Flow and supersonic flow over the BFS. The effect of grid refinement and the contribution of different SGS models on the mean flow properties and the turbulence quantities are analysed. To the authors' best knowledge, such a study does not exist in open literature and it will inspire further studies involving LES of flow through complex geometries with minimum computational cost. 

\section{Computational Methodology}
\subsection{Large Eddy Simulation} 
\quad Large Eddy Simulation involves the application of a filter function on the governing equations to yield filtered governing equations as presented below:\\
Continuity Equation:\\
\begin{equation}
\frac{\partial \overline{\rho}}{\partial t}+\frac{\partial ( \overline{\rho}\widetilde{u_i})}{\partial x_j}=0
\end{equation}
Momentum Equation:\\

\begin{equation}
\frac{\partial ( \overline{\rho}\widetilde{u_i})}{\partial t}+\frac{\partial ( \overline{\rho}\widetilde{u_i}\widetilde{u_j})}{\partial x_j}=
-\frac{\partial \overline{p}}{\partial x_i}+\frac{\partial }{\partial x_j}\widetilde{\sigma_{ij}}-\frac{\partial }{\partial x_j}\tau_{ij}
\end{equation}
Energy Equation:\\

\begin{equation}
\frac{\partial  \overline{E}}{\partial t}+\frac{\partial }{\partial x_i}(\widetilde{u_i}(\overline{E}+\overline{p}))=
-\frac{\partial }{\partial x_i}q_i+\frac{\partial }{\partial x_j}(\widetilde{\sigma_{ij}}\widetilde{u_i})
-\frac{\partial }{\partial x_i}(\overline{Eu_i} - \overline{E}\widetilde{u_i} + \overline{pu_i} - \overline{p}\widetilde{u_i})
\end{equation}
where,
\begin{equation}
  \begin{rcases}
\widetilde{\sigma_{ij}}=\overline{\mu}\left(\frac{\partial  \widetilde{u_i}}{\partial x_j}+\frac{\partial  \widetilde{u_j}}{\partial x_i}-\frac{2}{3}\delta_{ij}\frac{\partial  \widetilde{u_k}}{\partial x_k}\right)\\
q_i=-\overline{\kappa}\frac{\partial \widetilde{T}}{\partial x_i}\\
E = \rho\frac{T}{\gamma} + \frac{\rho u_iu_i}{2}\\
\tau_{ij}=\overline{\rho}(\widetilde{u_iu_j}-\widetilde{u_i}\widetilde{u_j})
   \end{rcases}
  \label{formulas}
\end{equation}

\quad The symbols $\mathtt{\sim}$ and \textendash\hspace{0.1 cm}refer to the filtered and Favre-filtered quantities respectively. E is the total energy per unit volume, $\mathtt{\gamma}$ is the ratio of specific heats, $\mathtt{\rho}$ is the density and $\mathtt{u_i}$ is the velocity vector. The fluid properties $\mathtt{\mu}$ and $\mathtt{\kappa}$ are the molecular viscosity and thermal conductivity, respectively. $\mathtt{\tau_{ij}}$ is the SGS stress tensor.

\subsection{Subgrid Scale Modeling}
\quad The application of filter function to the momentum equation gives rise to an SGS stress tensor, $\mathtt{\tau_{ij}}$. This stress tensor is modeled with the help of an SGS model using an eddy viscosity assumption as:\\
\begin{equation}
\tau_{ij} - \frac{1}{3}\tau_{kk}\delta_{ij} = -2\mu_{sgs}(\widetilde{S_{ij}} - \frac{1}{3}\widetilde{S_{kk}}\delta_{ij})
\end{equation}
where $\widetilde{S_{ij}} = \frac{1}{2}\left (\frac{\partial \widetilde{u_i}}{\partial x_j} + \frac{\partial \widetilde{u_j}}{\partial x_i}\right )$ is the strain rate tensor. The eddy viscosity is related to the filter width $\Delta$ and the strain rate magnitude $\widetilde{|S|} = \sqrt{2\widetilde{S_{ij}}\widetilde{S_{ij}}}$ as:\\
\begin{equation}
\mu_{sgs}=\overline{\rho}(C_s\Delta)^2\widetilde{|S|} 
\end{equation}
where $C_s$ is the Model constant.
In the Smagorinsky model, the value of $C_s$ is determined before the simulation and is kept fixed. The Dynamic Smagorinsky Model dynamically calculates $C_s$ based on the information from the flow.

\subsubsection{Dynamic Smagorinsky Model}
\quad The dynamic calculation of the Smagorinsky constant involves the application of a test filter denoted by $\hat{(.)}$  which is twice the size of the grid filter. This is achieved by utilizing the Germano identity
$L_{ij} = T_{ij} - \widehat{\tau_{ij}}$ \cite{germano1991dynamic}. The Leonard stresses \cite{pope2001turbulent} also termed as the resolved turbulent stresses are represented by $L_{ij} = \widehat{\overline{\rho u_i} \hspace{0.09 cm}\overline{\rho u_j}}/\overline{\rho}$ and the stresses due to the application of the test filter are represented by $T_{ij}=\widehat{\overline{\rho}}\hspace{0.09 cm} \widehat{\widetilde{u_i u_j}} - \widehat{\overline{\rho}}\hspace{0.09 cm} \widehat{\widetilde{u_i}} \widehat{\widetilde{u_j}}$. By substituting the Germano identity in Eq. 5 and using the Germano-Lilly procedure, the value of the Smagorinsky constant can be evaluated at every point for each time step as:
\begin{equation}
C_s^2=\frac{\left<L_{ij}M_{ij}\right>}{\left<M_{kl}M_{kl}\right>}
\end{equation}
where $\alpha_{ij}=-2\overline{\Delta}^2\overline{\rho}\widetilde{|S|} (\widetilde{S_{ij}} - \frac{1}{3}\widetilde{S_{kk}}\delta_{ij})$,
$\beta_{ij}=-2\widehat{\Delta}^2\widehat{\overline{\rho}}\widehat{\widetilde{|S|}} (\widehat{\widetilde{S_{ij}}} - \frac{1}{3}\widehat{\widetilde{S_{kk}}}\delta_{ij})$ and
$M_{ij}=\beta_{ij}-\widehat{\alpha_{ij}}$

\subsubsection{WALE Model}
\quad The WALE model, as mentioned before, overcomes the limitations of Smagorinsky model by employing a different velocity scale. While the latter uses $\widetilde{|S|} $ as the velocity scale, the former uses the scale $\frac{(S_{ij}^d S_{ij}^d )^{3/2}}{(\widetilde{S_{ij}} \widetilde{S_{ij}})^{5/2} + (S_{ij}^d S_{ij}^d )^{5/4}}$ and calculates the eddy viscosity as:\\ 

\begin{equation}
\mu_{sgs}=\overline{\rho}(C_w\Delta)^2 \frac{(S_{ij}^d S_{ij}^d )^{3/2}}{(\widetilde{S_{ij}} \widetilde{S_{ij}})^{5/2} + (S_{ij}^d S_{ij}^d )^{5/4}}
\end{equation}
where $S_{ij}^d = \frac{1}{2}({\widetilde{g_{ij}}^2} + {\widetilde{g_{ji}}^2})$, $\widetilde{g_{ij}} = \frac{\partial \widetilde{u_i}}{\partial x_j}$ and $C_w$ is the WALE constant whose value is taken as 0.5.

\subsubsection{Dynamic WALE Model}
\quad Similar to the Smagorinsky constant, the WALE constant cannot be universal especially for complex geometries where a fixed value might incorrectly predict the eddy viscosity.  Therefore, Germano-Lilly procedure was applied to the WALE model by Toda et al. \cite{toda2010dynamic} which resulted in a Dynamic WALE Model. However they observed that the resulting model over estimated the eddy viscosity due to high value of the WALE constant near the wall. One can conclude that the models which have correct near wall behavior are sensitive to the filtering procedure. Since the molecular viscosity is dominant in the near wall region, any model with correct wall behavior does not need to calculate the model constant dynamically. Considering this fact, they came up with a Shear and Vortex Sensor (SVS), a quantity used to detect a near wall. The SVS is given by $\frac{(S_{ij}^d S_{ij}^d )^{3/2}}{(\widetilde{S_{ij}} \widetilde{S_{ij}})^{3} + (S_{ij}^d S_{ij}^d )^{3/2}}$ which is 0 for pure shear flows and 1 for pure rotating flows. The dynamic procedure is applied for SVS values greater than 0.09 and a fixed value of WALE constant(=0.5) is used for SVS less than 0.09.

\subsection{Computational Details}
\quad Figures 1-3 depict the computational grids for the three cases studied herein, where BFS-I represents the subsonic BFS geometry, BFS-II denotes the supersonic BFS geometry and CSF designates the confined swirling flow geometry. 

\begin{figure}[H]{}
	\centering
  \includegraphics[width=\linewidth,trim=10 10 10 10,clip]{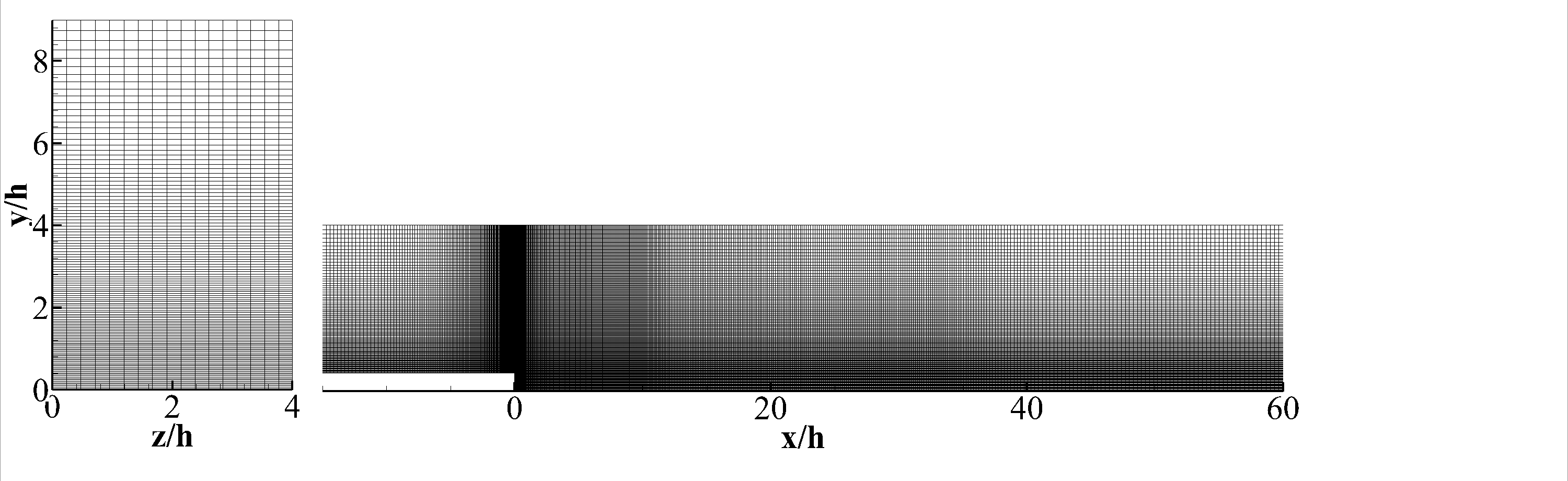}

\caption{The schematic of mid-plane grid distribution for BFS-I}\label{fig:1}
\end{figure}

\begin{figure}[H]{}
	\centering
	\includegraphics[width=\linewidth,trim=5 5 5 5,clip]{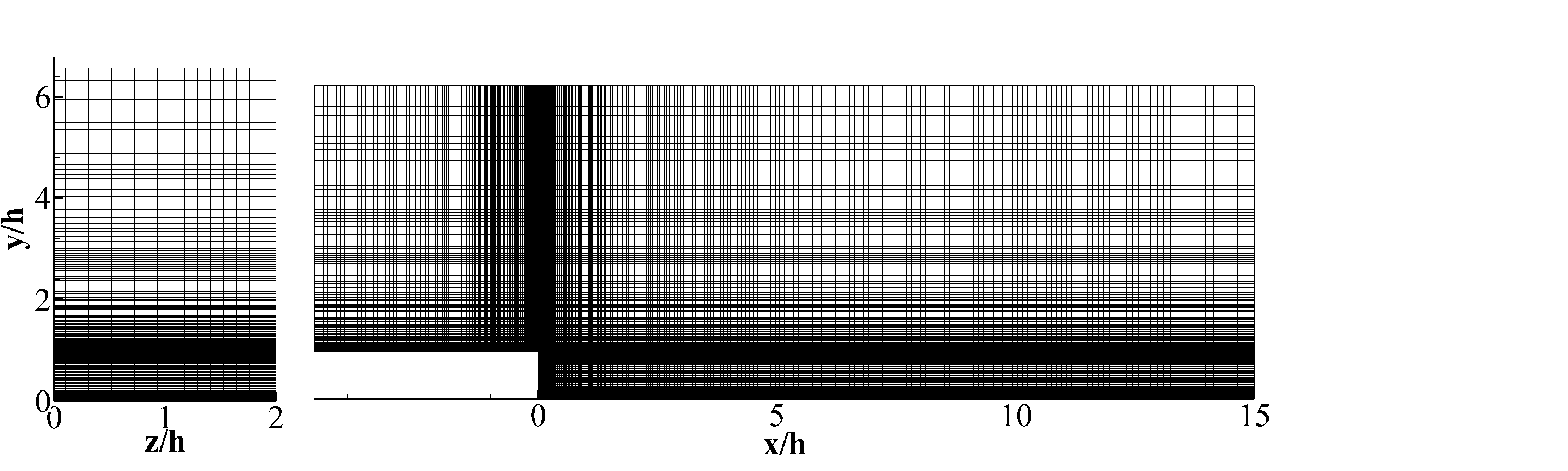}
	
	\caption{The schematic of mid-plane grid distribution for BFS-II(supersonic case)}\label{fig:2}
\end{figure}

as CSF hereafter. For each case, the generated grid comprises of multiblock grid structure which is successively stretched along the wall normal direction as well as in the axial direction such that the aspect ratio of the cells in the regions of interest is not significant. We consider two sets of grids for each geometry based on the non-dimensional wall coordinate ($y^+$).The geometry extends upto 60 step heights and 15 step heights after the step for BFS-I and BFS-II, respectively in the downstream direction. For CSF, the geometry extends upto 25D in the axial direction, where, D is the diameter of the inlet. The diameter of the tube after the expansion corner is two times the inlet diameter.The details of the geometry are provided in Table \ref{tab:1}. The characteristic length for BFS-I and BFS-II is the step height while for CSF, it is the inlet diameter.

\begin{figure}[H]{}
	\centering
	\includegraphics[width=\linewidth,trim=5 5 5 5,clip]{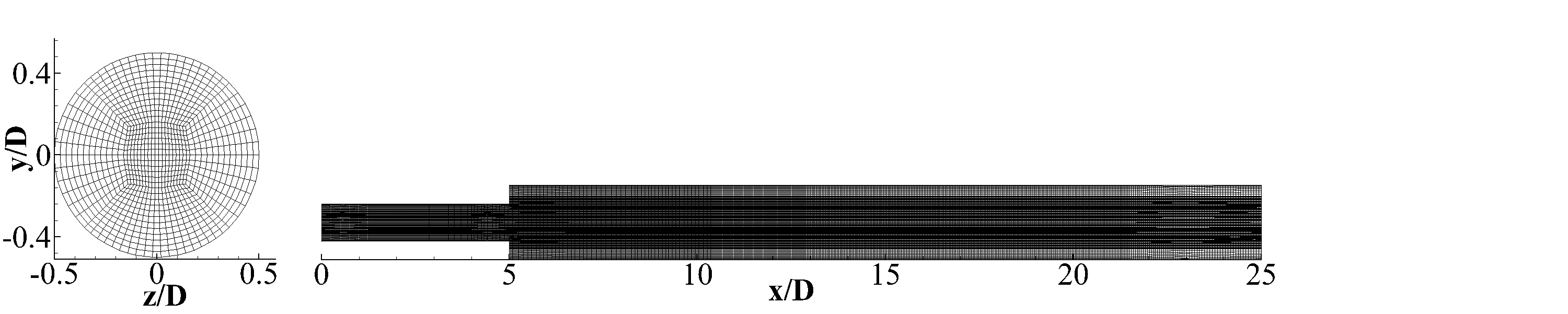}
	
	\caption{The grid distribution for CSF}\label{fig:3}
\end{figure}

\quad A power law inlet profile is imposed for the axial velocity for all the cases. In addition to this, a uniform swirling inlet (Swirl No. = 0.6) is provided for the CSF. The cylindrical walls of CSF and the bottom walls of the BFS-I and BFS-II use No-slip boundary condition, while the top surfaces of BFS cases use freestream condition. The spanwise direction of the BFS cases invokes the periodic boundary condition, while the outlet face for all the cases use convective boundary condition. 

\begin{table}[H]
\centering
\caption{Details of different flow configurations}\label{tab:1}
\begin{tabularx}{\linewidth}{| X | X | p{2.4cm} |p{2.4cm}|p{2.4cm}|}
\hline
Geometry & Reynolds No. & Characteristic Length(m) & Grid Points(Coarse) & Grid Points (Fine) \\
\hline
BFS-I & 37200 & 0.0127 & 1.3M ($y^+=$70) & 3.0M ($y^+=$20 )\\
\hline
BFS-II & 100000 & 0.0032 & 1.5M ($y^+=$40) & 2.5M ($y^+=$20) \\
\hline
CSF & 30000 & 0.0508 & 1.8M ($y^+=$40) & 3.5M ($y^+=$20) \\
\hline
\end{tabularx}
\end{table}

\subsection{Numerical Method}
\quad All the three SGS models have been incorporated in an in-house code which uses a density based solver in a FVM framework to solve the governing equations. Low Mach number preconditioning \cite{weiss1995preconditioning} is used to solve for the incompressible flow regime. A second order Low Diffusion Flux-Splitting Scheme \cite{edwards1997low} has been used to discretize the convective fluxes.The Diffusive terms are calculated using second order central difference scheme. The temporal terms use an implicit second order discretization. The solver had been extensively used for a large number of studies including both reacting and non-reacting flows \cite{das2015numerical,das2016numerical,de2009large,de2009largee,de2009experimental,de2012parametric,de2012dynamics}.

\section{Results and Discussion}
\begin{figure}[H]
	\begin{subfigure}[b]{0.3\linewidth}
		\includegraphics[width=4cm,height=5cm,trim=50 50 50 50,clip]{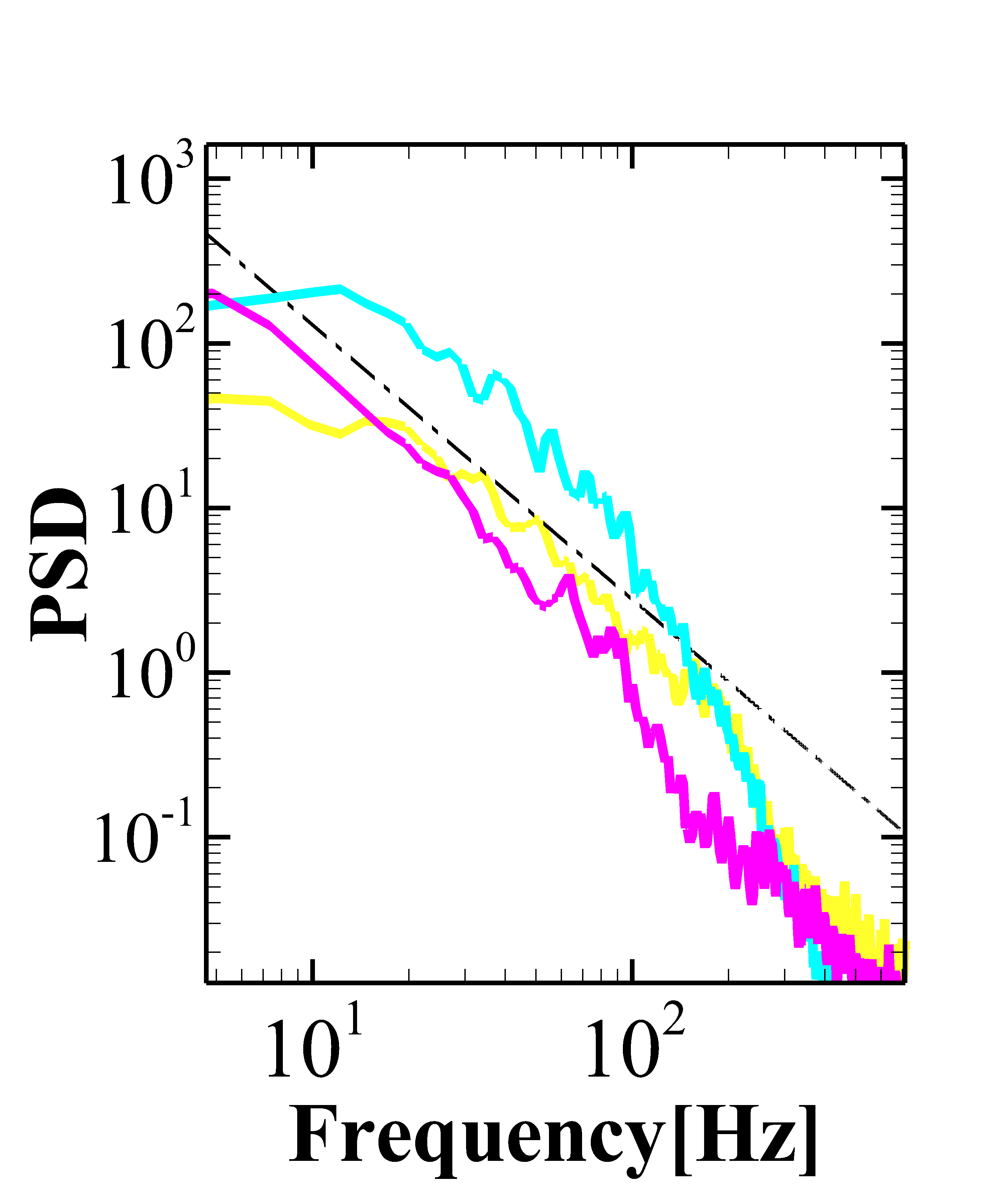}
		\caption{}\label{fig:4a}
	\end{subfigure}
	\begin{subfigure}[b]{0.3\linewidth}
		\includegraphics[width=4cm,height=5cm,trim=50 50 50 50 ,clip]{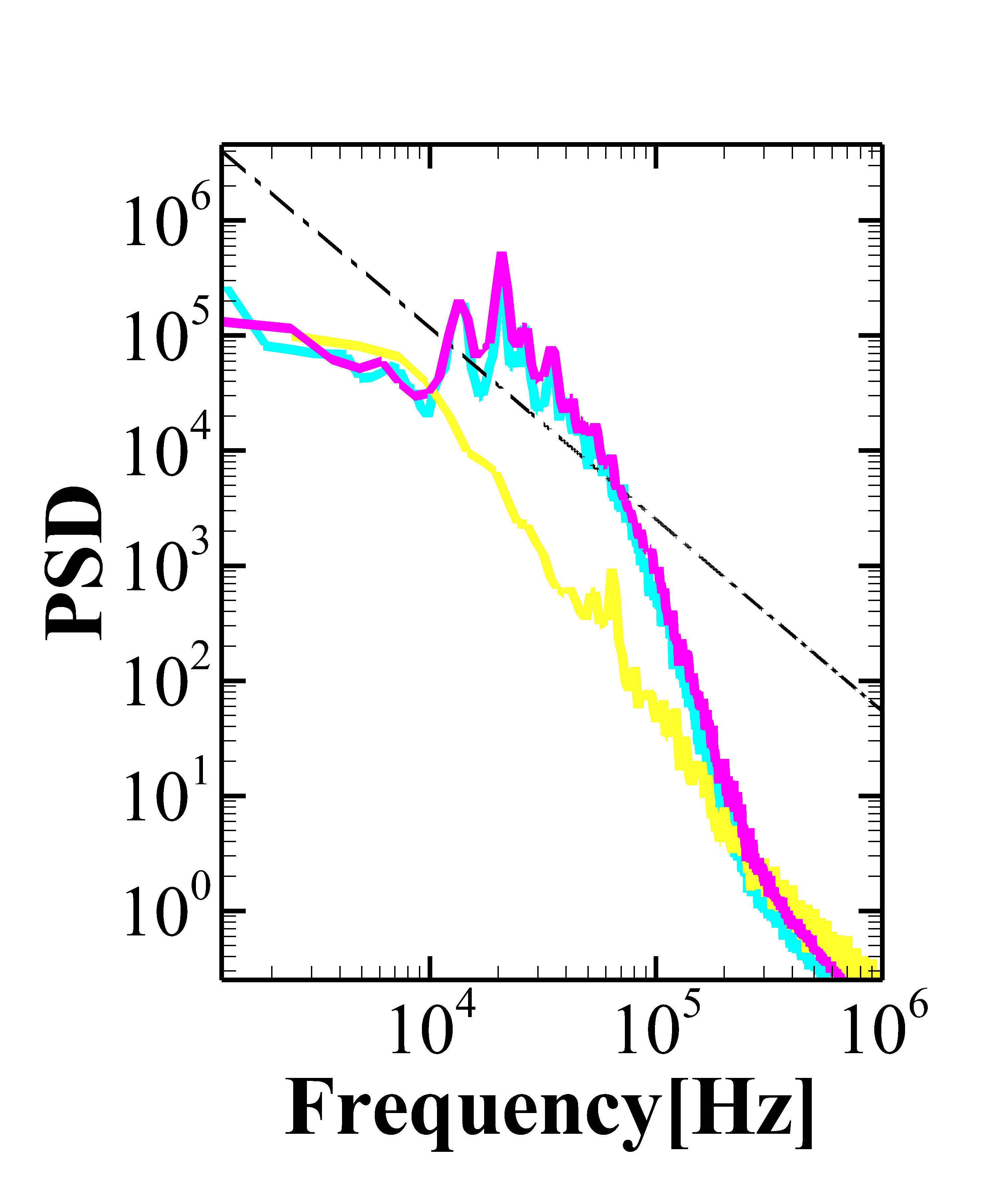}
		\caption{}\label{fig:4b}
	\end{subfigure}
	\begin{subfigure}[b]{0.3\linewidth}
		\includegraphics[width=4cm,height=5cm,trim=50 50 50 50,clip]{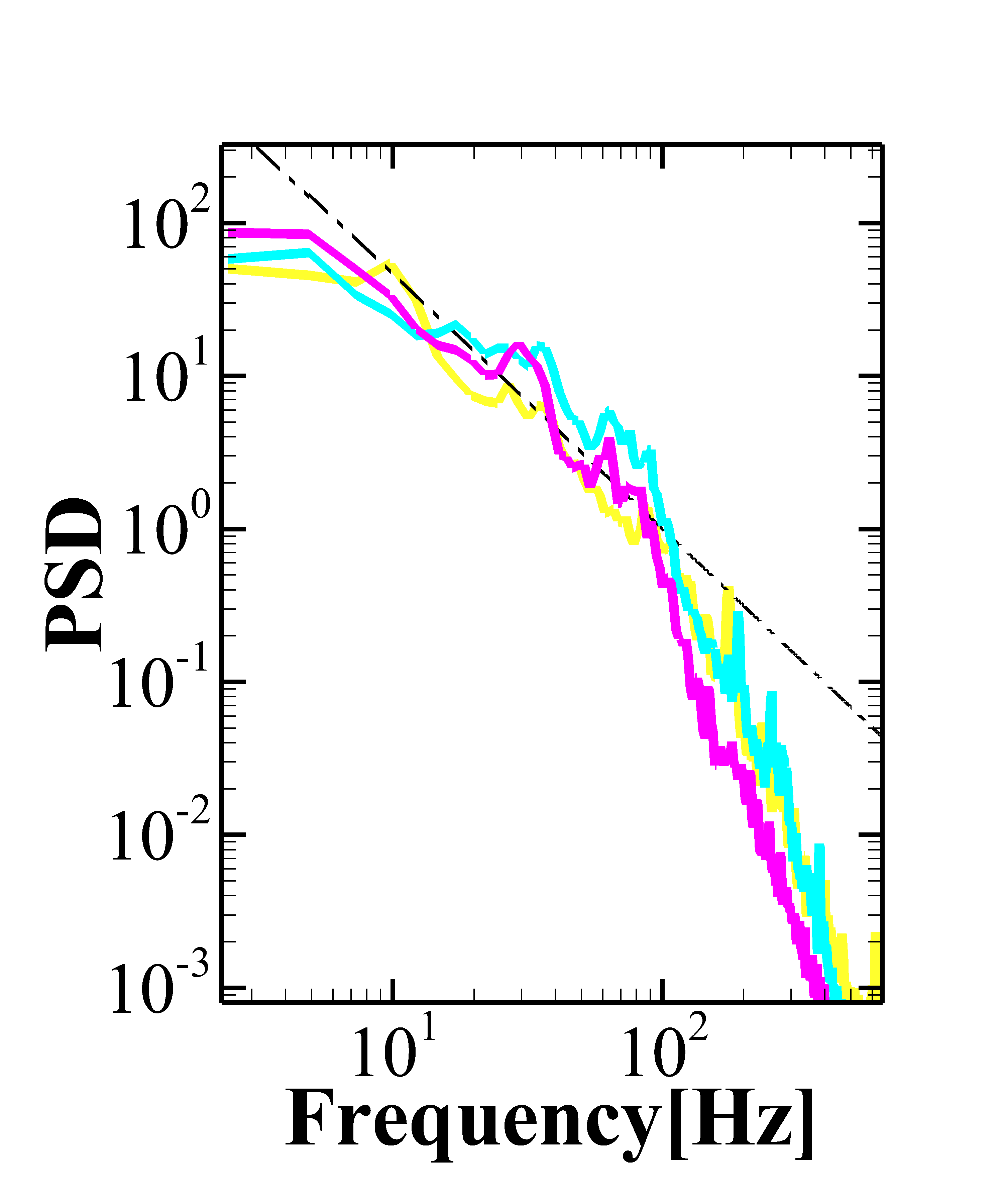}
		\caption{}\label{fig:4c}
	\end{subfigure}
	\caption{Energy Spectrum }\label{fig:4}
\end{figure}

\quad As mentioned earlier, there are specific quality parameters to check the suitability of the grid for LES. One such parameter is the Energy Spectrum. The velocity spectra for all the grids have been obtained. The difference between the coarse grid spectra and the fine grid spectra is not very significant. Hence, only the spectra for the fine grids for all the models for BFS-I, BFS-II and CSF are presented in the Fig. \ref{fig:4a}, Fig. \ref{fig:4b} and Fig. \ref{fig:4c}, respectively. Unless specified in the figure, Table \ref{tab:2} provides the legend for all the figures. The black dotted line represents the -5/3 slope of the Kolmogorov Spectrum. It can be observed that a sufficient inertial range is present for all the cases thus establishing a sufficient resolution for the present LES. In addition to the energy spectra, the quality of the grid has also been tested using the criterion proposed 

\begin{figure}[H]
	\centering
	\includegraphics[width=20cm,height=8cm,keepaspectratio, trim=500 50 500 100, clip]{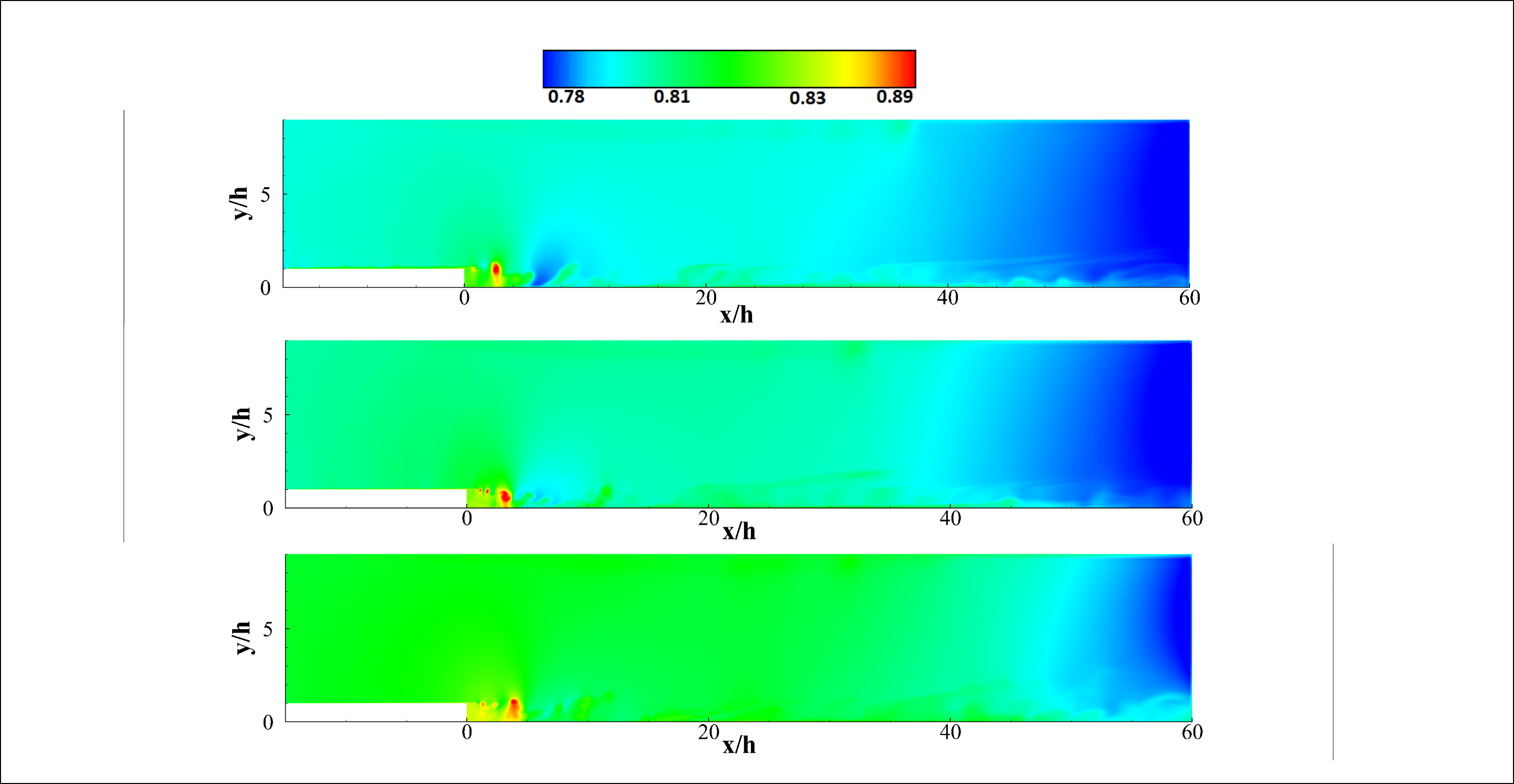}
\caption{$LES_{IQ}$ for BFS-I}\label{fig:5}
\end{figure}

\begin{figure}[H]
	\centering
	\includegraphics[width=20cm,height=8cm,keepaspectratio, trim=500 150 500 100, clip]{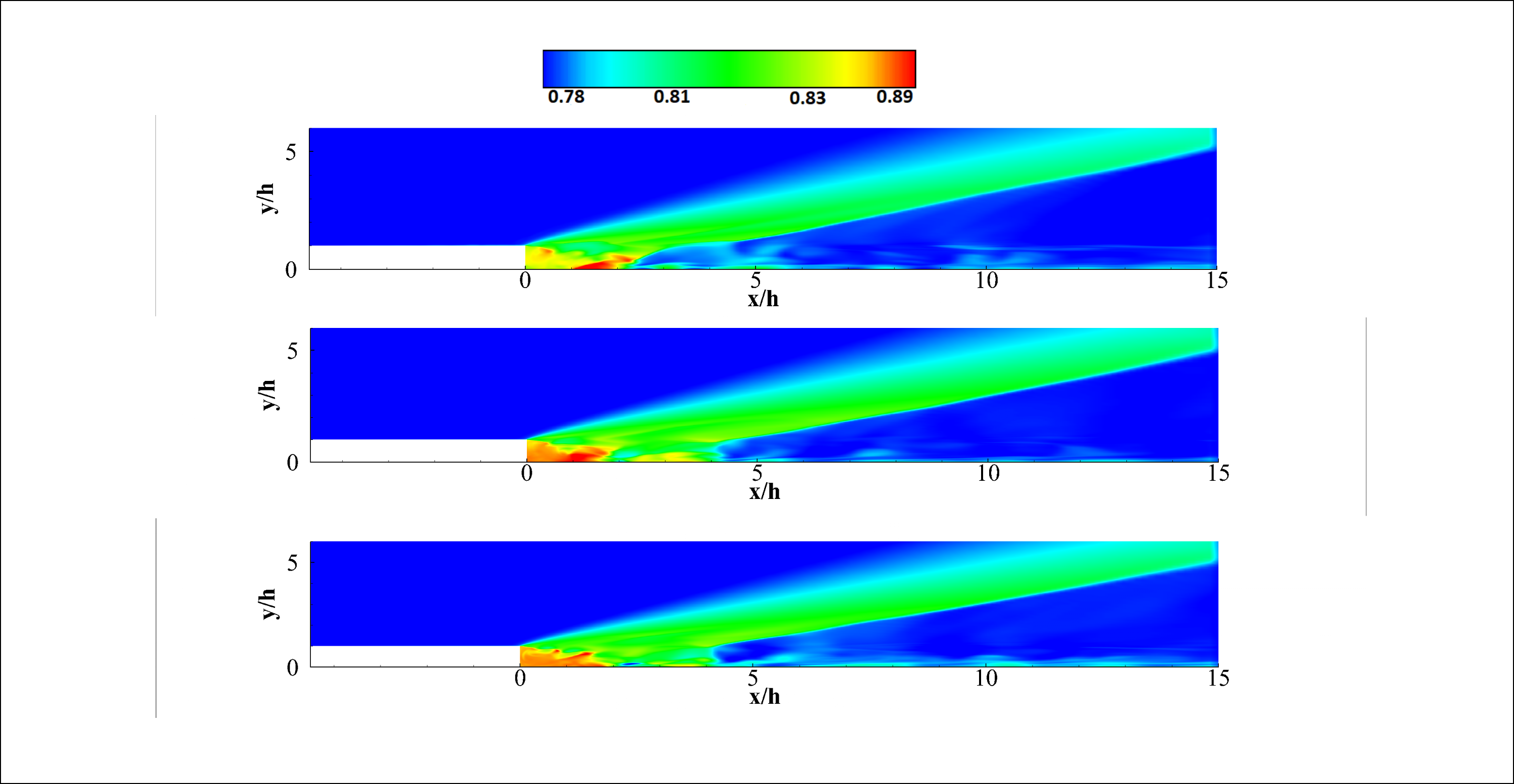}
	\caption{$LES_{IQ}$ for BFS-II}\label{fig:6}
\end{figure}

\begin{figure}[H]
	\centering
	\includegraphics[width=20cm,height=8cm,keepaspectratio, trim=500 200 500 100, clip]{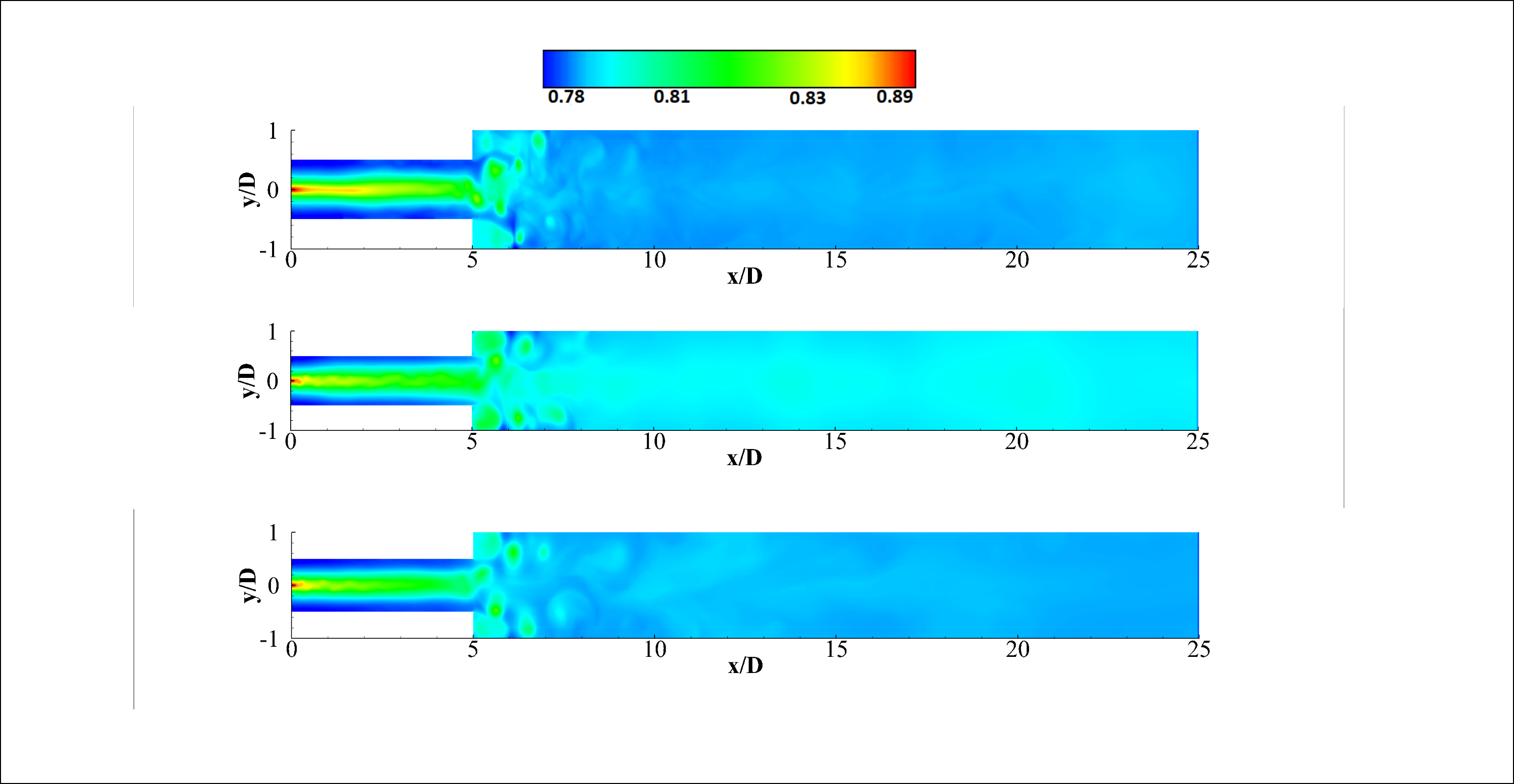}
	\caption{$LES_{IQ}$ for CSF}\label{fig:7}
\end{figure}

\begin{table}[H]
\centering
\caption{Legend for all the figures}\label{tab:2}
\begin{tabularx}{\linewidth}{| X | X |X|}
\hline
Model & Grid & line\\
\hline
Dynamic Smagorinsky & coarse & 
\begin{tikzpicture}
\draw[green,thick] (0,0) -- (1,0); 
\end{tikzpicture} \\
\hline
Dynamic Smagorinsky & fine& 
\begin{tikzpicture}
\draw[yellow,thick] (0,0) -- (1,0); 
\end{tikzpicture} \\
\hline
WALE & coarse& 
\begin{tikzpicture}
\draw[black,thick] (0,0) -- (1,0); 
\end{tikzpicture} \\
\hline
WALE & fine& 
\begin{tikzpicture}
\draw[cyan,thick] (0,0) -- (1,0); 
\end{tikzpicture} \\
\hline
Dynamic WALE & coarse& 
\begin{tikzpicture}
\draw[blue,thick] (0,0) -- (1,0); 
\end{tikzpicture} \\
\hline
Dynamic WALE & fine& 
\begin{tikzpicture}
\draw[magenta,thick] (0,0) -- (1,0); 
\end{tikzpicture} \\
\hline
\end{tabularx}
\end{table}

by Celik et al. \cite{celik2005index}. The parameter $LES_{IQ}$ is related to the eddy viscosity and kinematic viscosity as:
\begin{equation}
LES_{IQ} = \frac{1}{1+0.05(\frac{\nu+\nu_t}{\nu})^{0.53}}
\end{equation}
\quad A value of $LES_{IQ}$ greater than 0.75 is considered to be a well resolved LES. The $LES_{IQ} $ for the fine grids for all the models is presented in the  Fig. \ref{fig:5}, Fig. \ref{fig:6} and Fig.\ref{fig:7}. It can be observed that for all the grids the value of the parameter is well above 0.75. Thus, the LES for the present cases has an adequate resolution and is fit for further analysis. The results for different cases are discussed separately in the subsequent sections.

\subsection{BFS-I}
\begin{figure}[H]
\centering
\includegraphics[width=12cm,height=8cm, trim=10 10 10 10, clip]{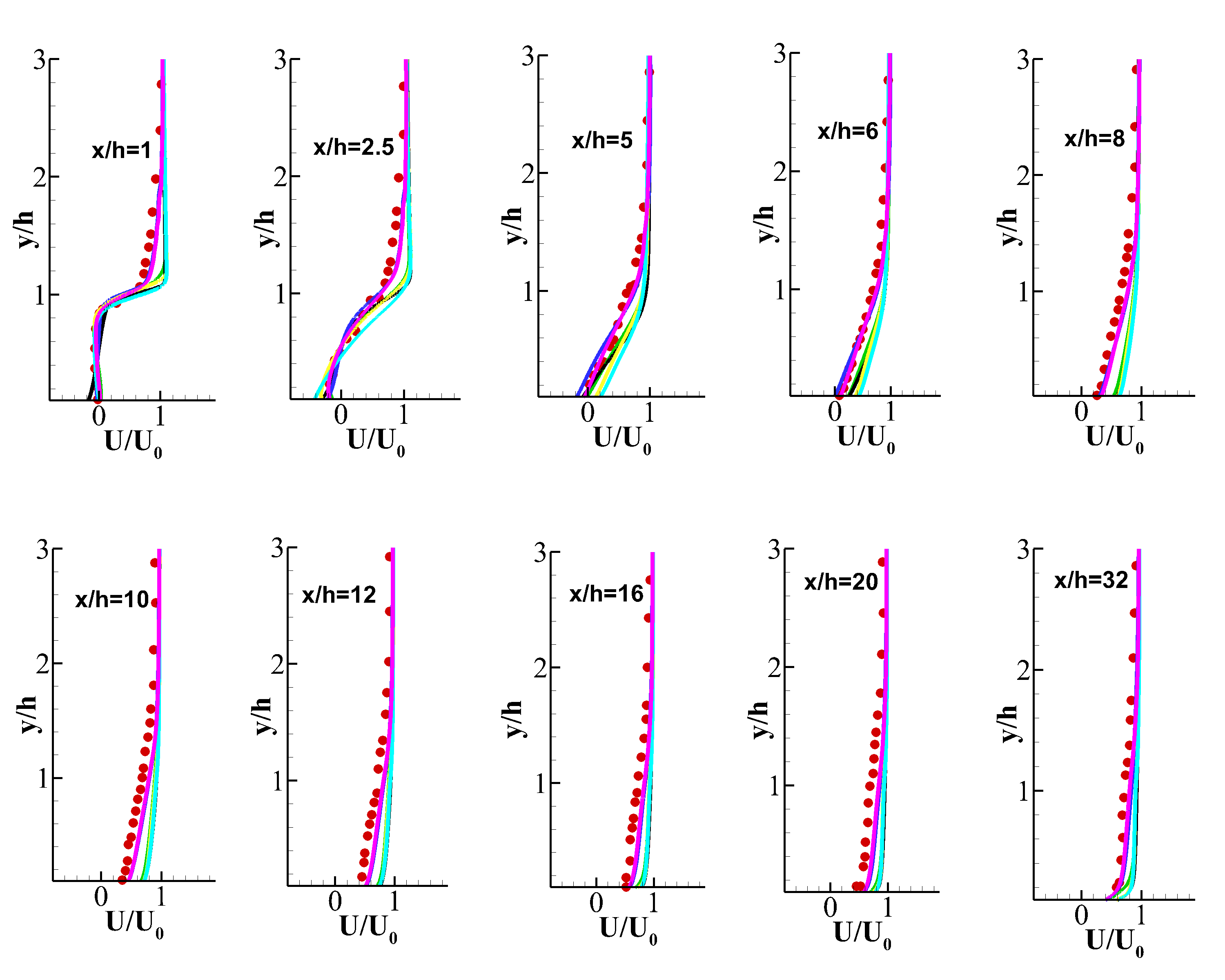}
\caption{Mean Streamwise Velocity profiles for BFS-I}\label{fig:8}
\end{figure}

\quad The flow over a BFS is an exciting problem \cite{eaton1981review,kuehn1980effects,narayanan1974similarities} which proceeds with the flow separation at the step corner followed by reattachment at the base wall at some downstream location. This gives rise to a separation bubble at the step corner. The reattachment length is an important parameter which depends on many factors that include the contraction ratio, and upstream boundary layer thickness \cite{armaly1983experimental, driver1985features}. The results from the present LES simulations are validated against the experimental findings of Driver and Seegmiller \cite{driver1985features}. Figure \ref{fig:8} presents the mean velocity profiles along different axial locations. The velocity profiles for coarse and fine grid for the Dynamic WALE model at $x/h=1$ show that the separation is predicted well by them. For the other two models, the mean streamwise velocity at $y/h=1$ is over-predicted. Table \ref{tab:3} shows the reattachment length for all the cases. Again, a quick observation shows that the Dynamic WALE model predicts the reattachment length close to the experimental values for both the grid resolutions.

\begin{figure}[H]
\centering
\includegraphics[width=12cm,height=8cm, trim=10 10 10 10, clip]{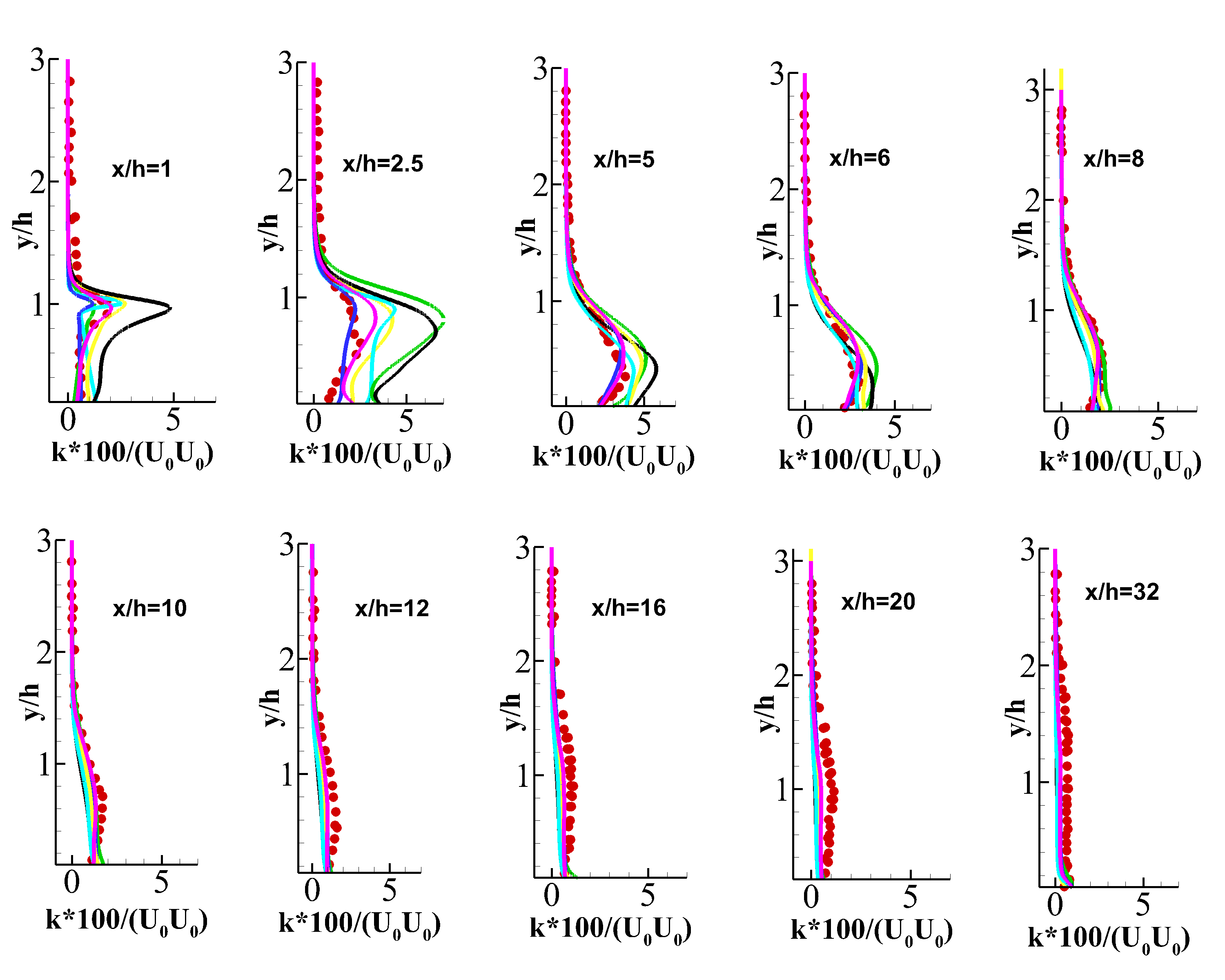}
\caption{Turbulent Kinetic Energy profiles for BFS-I}\label{fig:9}
\end{figure}

\begin{figure}[H]
\centering
\includegraphics[width=12cm,height=8cm, trim=10 10 10 10, clip]{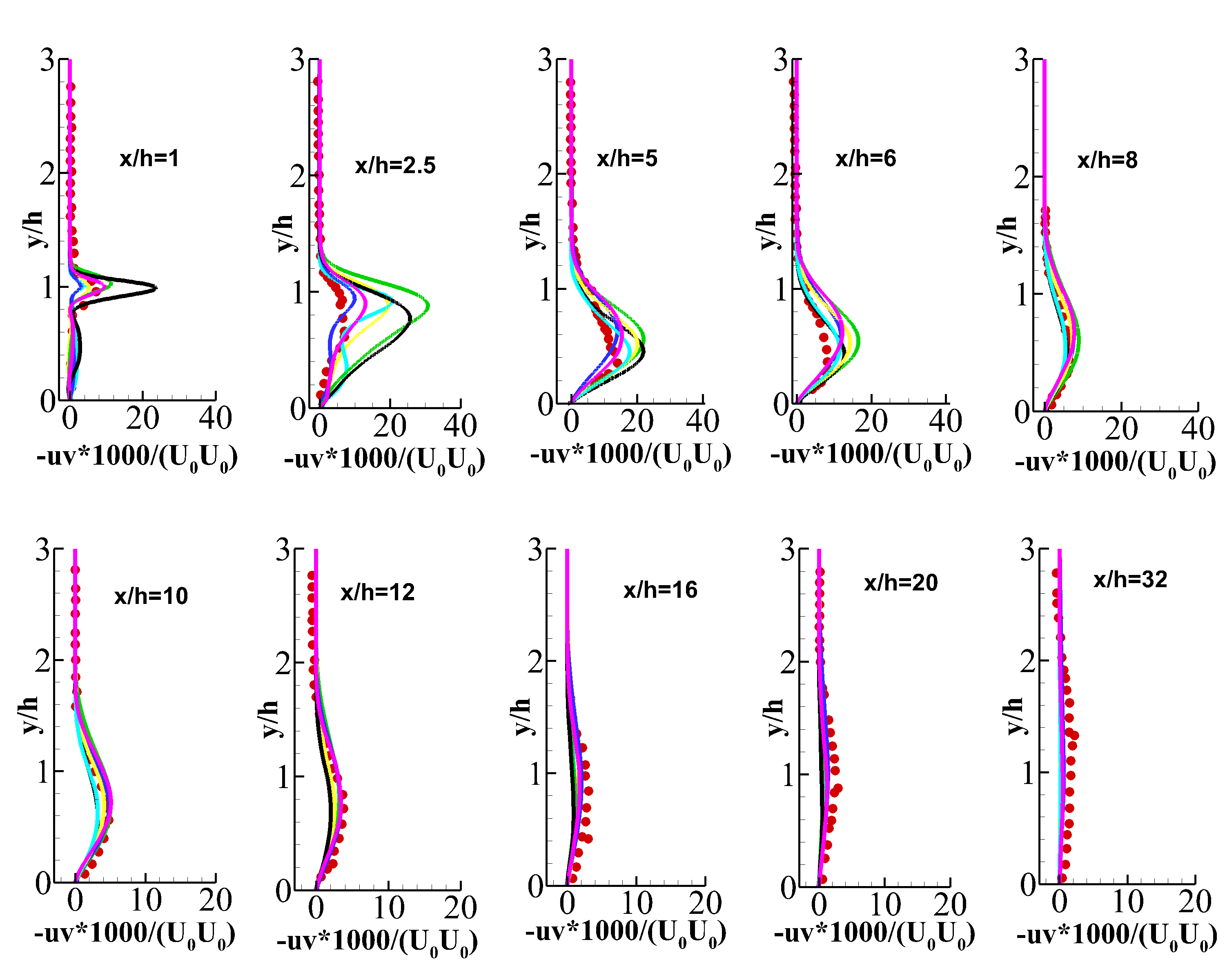}
\caption{Turbulent Shear Stress profiles for BFS-I}\label{fig:10}
\end{figure}

\quad The velocity profiles for the WALE and the Dynamic Smagorinsky model are flatter near the experimental reattachment location ($x/h \approx 6$) due to incorrect prediction of the reattachment length by these models. Figure \ref{fig:9} depicts the turbulent kinetic energy at different axial locations. The peak in the kinetic energy profile occurs near the step corner following the formation of the shear layer. It increases steadily upto approximately $x/h=5$. Since the Dynamic WALE predicts the separation well, the kinetic energy profile at $x/h=1$ follows the experimental trends closely. The WALE model for the coarse grid over-predicts the kinetic energy at this location whereas, the values predicted by the Dynamic Smagorinsky model for both the grids are also close to the experimental values. The turbulent shear stress presented in the Fig. \ref{fig:10} also shows similar behavior as the turbulent kinetic energy profile. The WALE model again over predicts the value of the shear stress at $x/h=1$. The Dynamic Smagorinsky as well as the WALE model over predicts the value of turbulent kinetic energy as well as the turbulent shear stress upto $x/h=5$ after which the kinetic energy and the shear stress starts to decay and the prediction of all the models are in accordance with the experimental observations.

\subsubsection{Near Wall Eddy Viscosity profile}
\quad The expansion which the flow undergoes at the step corner induces an adverse pressure gradient in the flow. It is a fact that the developing boundary layer is similar to the boundary layer over a flat plate, although it significantly deviates from the law of the wall \cite{nagano1993effects} due to the effect of this adverse pressure gradient. This behavior is observed even after many step heights downstream of the reattachment and it depends on many geometric as well as flow parameters. Nevertheless, the near wall behavior of the eddy viscosity predicted by all the models can be observed on this adverse pressure gradient boundary layer which starts to develop after the reattachment point.

\begin{table}[H]
\centering
\caption{Reattachment lengths($x/h$) for different models}\label{tab:3}
\begin{tabularx}{\linewidth}{| X | X | X |}
\hline
Experiment/Model & Reattachment Length & \% error(absolute) \\
\hline
Experiment & 6.26 & 0 \\
\hline
Dynamic Smagorinsky (coarse) & 5.30 & 15.33 \\ 
\hline
Dynamic Smagorinsky(fine) & 5.40 & 13.73 \\
\hline
WALE (coarse) & 5.60 & 10.54 \\
\hline
WALE (fine) & 5.70 & 8.94 \\
\hline
Dynamic WALE (coarse) & 6.30 & 0.63 \\ 
\hline
Dynamic WALE (fine) & 6.25 & 0.15 \\ 
\hline
\end{tabularx}
\end{table}

\begin{figure}[H]
\begin{subfigure}[b]{0.3\linewidth}
\includegraphics[width=4cm,height=5cm,trim=50 50 50 50,clip]{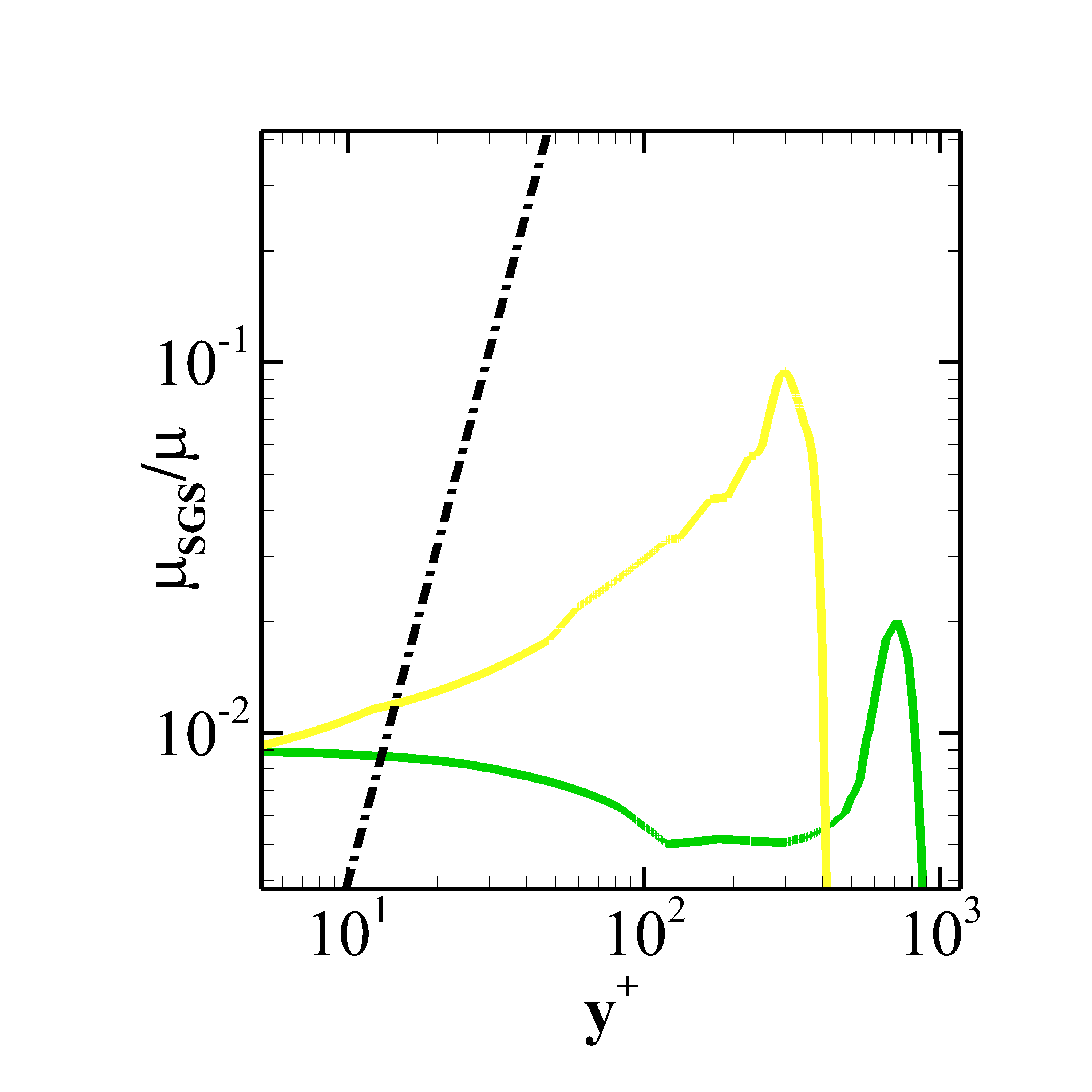}
\caption{}
\end{subfigure}
\begin{subfigure}[b]{0.3\linewidth}
\includegraphics[width=4cm,height=5cm,trim=50 50 50 50 ,clip]{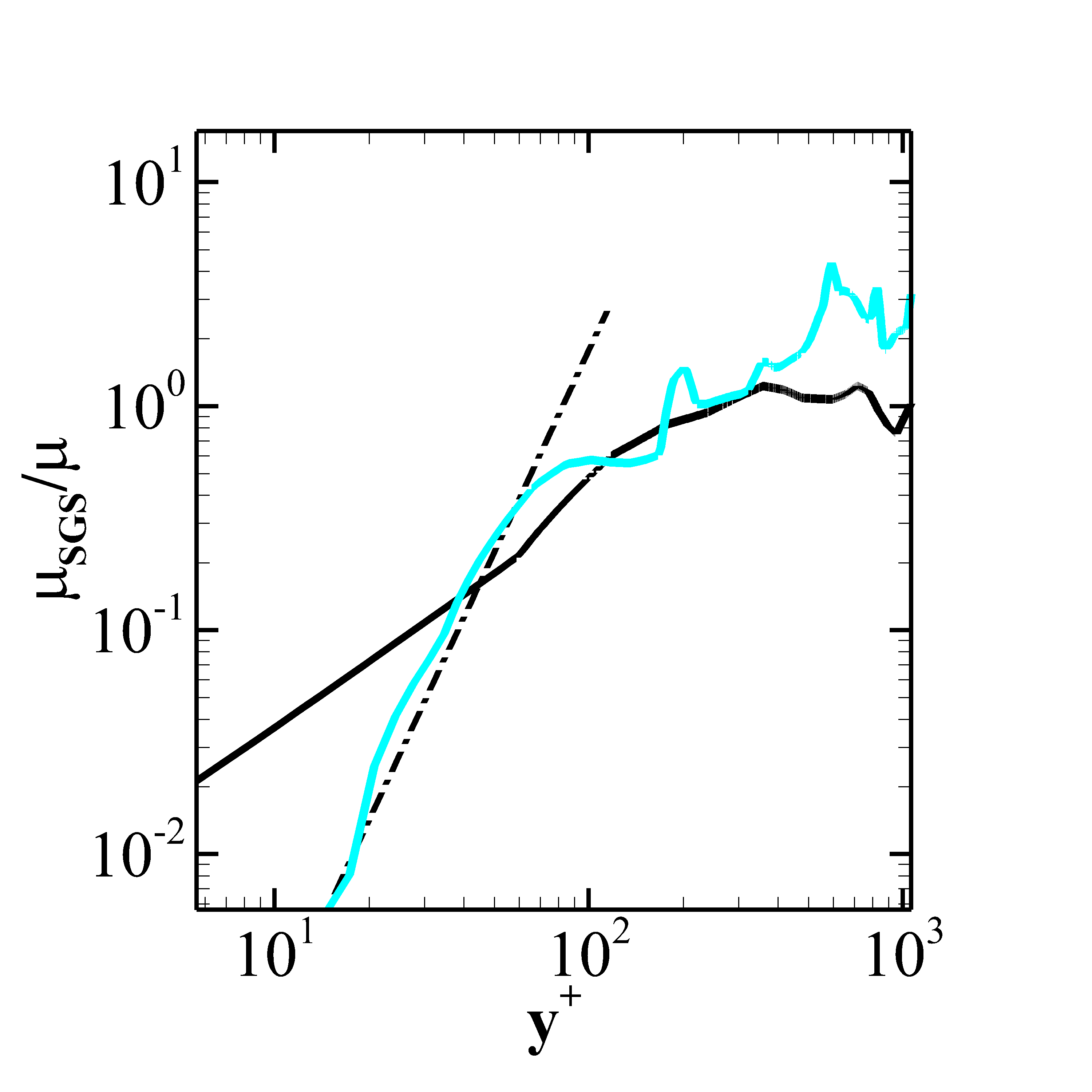}
\caption{}
\end{subfigure}
\begin{subfigure}[b]{0.3\linewidth}
  \includegraphics[width=4cm,height=5cm,trim=50 50 50 50,clip]{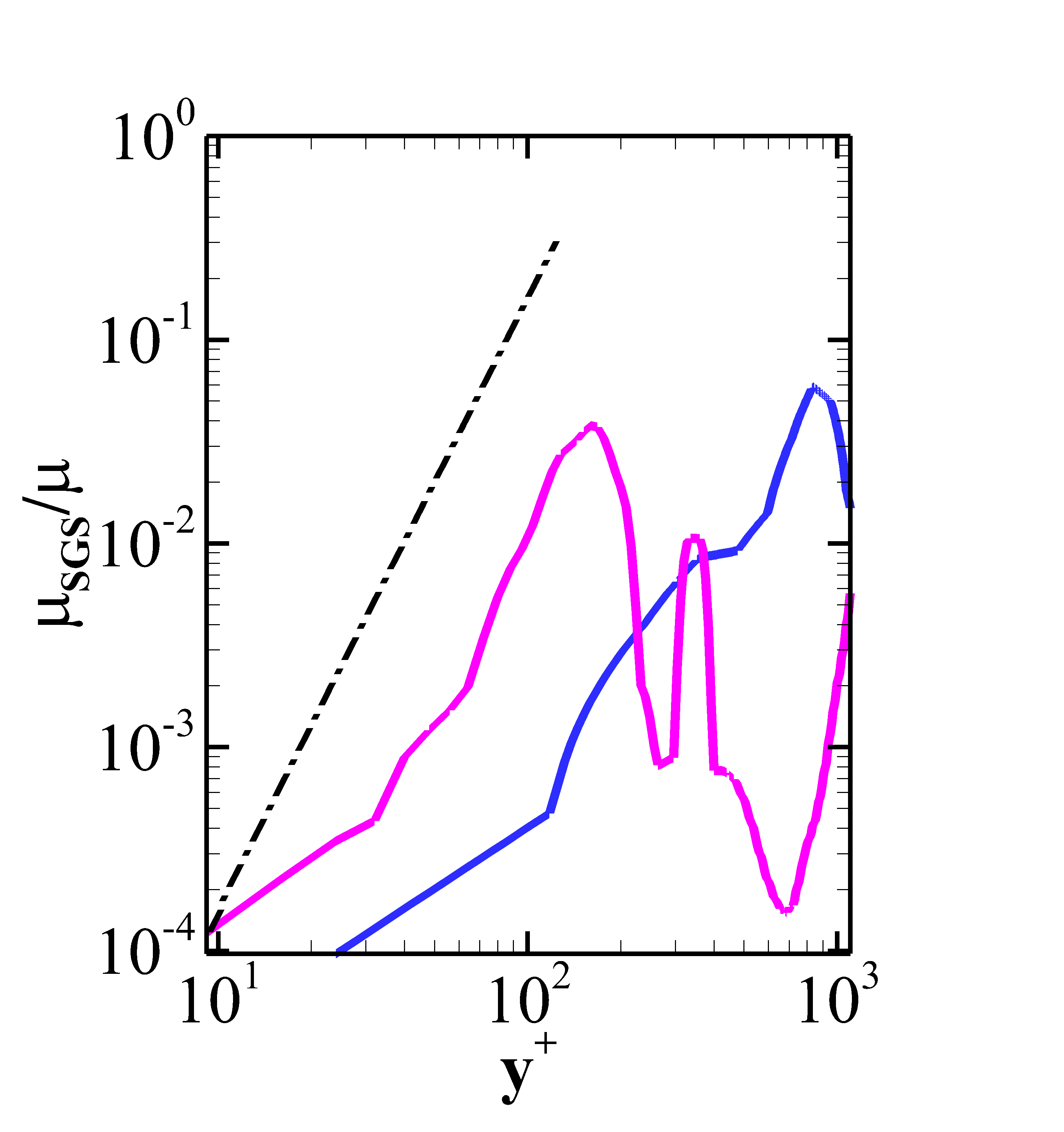}
  \caption{}
\end{subfigure}
\caption{Ratio of Eddy-Viscosity to Molecular Viscosity}\label{fig:11}
\end{figure}

\quad The eddy viscosity profiles are obtained at $x/h=40$ because it is sufficiently downstream of the reattachment point and the aspect ratio of the cells is moderate since the grid stretching is fairly reasonable in the axial direction. Figure \ref{fig:11} shows the ratio of eddy viscosity to the molecular viscosity, where the black dashed line represents the $y^{+^3}$ line. One can note that the eddy viscosity predicted by the fine grid WALE model follows a $y^{+^3}$ profile near the wall.

\quad The Dynamic Smagorinsky model, as expected, does not follow the desired profile near the wall region. The eddy viscosity is also high in the near wall region as compared to both the models. As discussed earlier, the Dynamic WALE model behaves like a WALE model in the vicinity of a wall. However, the near wall profile of the eddy viscosity predicted by the Dynamic WALE model is definitely not better or as good as the WALE model. This is attributed to the calculation of the Shear and Vortex Sensor(SVS). The applicability of the dynamic procedure in this model is restricted for SVS higher than 0.09 which was well established for a channel flow case\cite{toda2010dynamic}. The streamwise velocity for an adverse pressure gradient boundary layer lies below the universal log law \cite{le1997direct} . Therefore, the strain rates and hence the SVS  will have different profiles which requires the cut off condition for the application of the dynamic procedure to be different. Further, the shear and vortex sensor is a quantity which is not unique. The improvement of near wall behavior of the Dynamic WALE model may also be possible by considering a different SVS.

\quad Apart from comparing the mean velocity field and turbulent stresses to the experimental values, the performance of different SGS models can also tested by quantifying the dissipation provided by them. This exercise involves calculation of an SGS activity paramter which is discussed in the next section.
\subsubsection{SGS activity parameter}
\quad The SGS model provides dissipation in the LES framework. Therefore, to assess the performance of any SGS model, it is essential to quantify the dissipation provided by it. This is achieved with the help of an SGS activity parameter proposed by Geurts and Frohlich \cite{geurts2002framework}. The SGS activity paramter 's' is given by the following relation: \\
\begin{equation}
s=\frac{\epsilon_t}{\epsilon_t+\epsilon_m}
\end{equation}
where, $\epsilon_t$ is the total dissipation provided by the SGS model and $\epsilon_m$ is the molecular dissipation. The total dissipation provided by the SGS model is composed of dissipation due to the resolved fluctuations, $\epsilon_t'$ and the disspation due to the time averaged velocity field, $\epsilon_t^{mean}$ \cite{davidson2006transport, ben2017assessment}. The SGS dissipation due to the resolved fluctuating field is given by:
\begin{equation}
\epsilon_t = -2<\mu_{sgs}s_{ij}'s_{ij}'>
\end{equation}
where $s_{ij}' = \frac{1}{2}( \frac{\partial u_i'}{\partial x_j} + \frac{\partial u_j'}{\partial x_i} )$ is the fluctuating strain tensor. The SGS dissipation due to the time averaged mean field is given by:

\begin{equation}
\epsilon_t^{mean} = -2<\mu_{sgs}>S_{ij}S_{ij}
\end{equation}

where $S_{ij}=\frac{1}{2}( \frac{\partial \overline{u_i}}{\partial x_j} + \frac{\partial \overline{u_j}}{\partial x_i} )$  is the strain rate due to the mean velocity field. The symbol '$< >$' represents the averaged value.

\begin{figure}[H]
\begin{subfigure}[b]{0.3\linewidth}
\includegraphics[width=4cm,height=5cm,trim=50 50 50 50,clip]{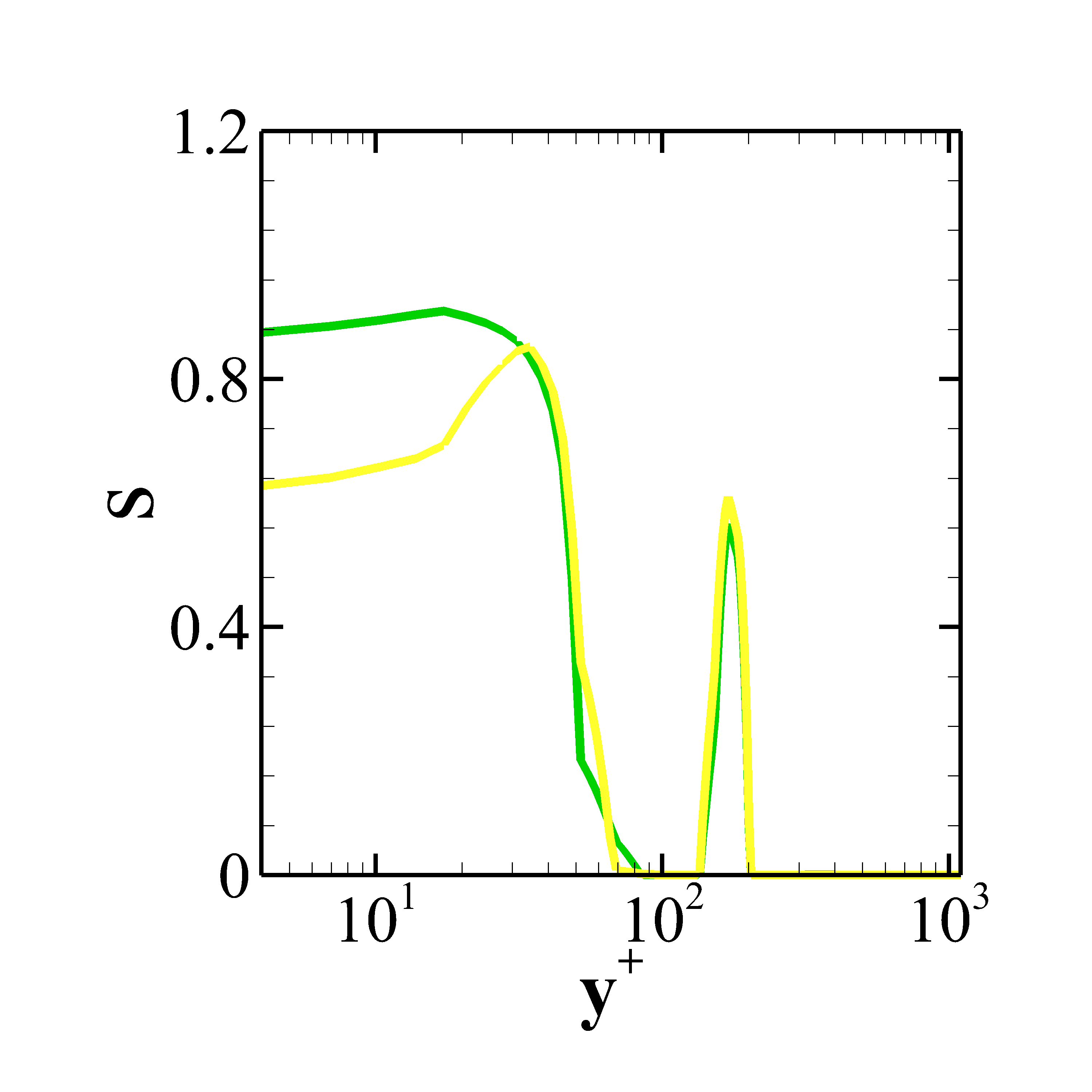}
\caption{}\label{fig:12a}
\end{subfigure}
\begin{subfigure}[b]{0.3\linewidth}
\includegraphics[width=4cm,height=5cm,trim=50 50 50 50 ,clip]{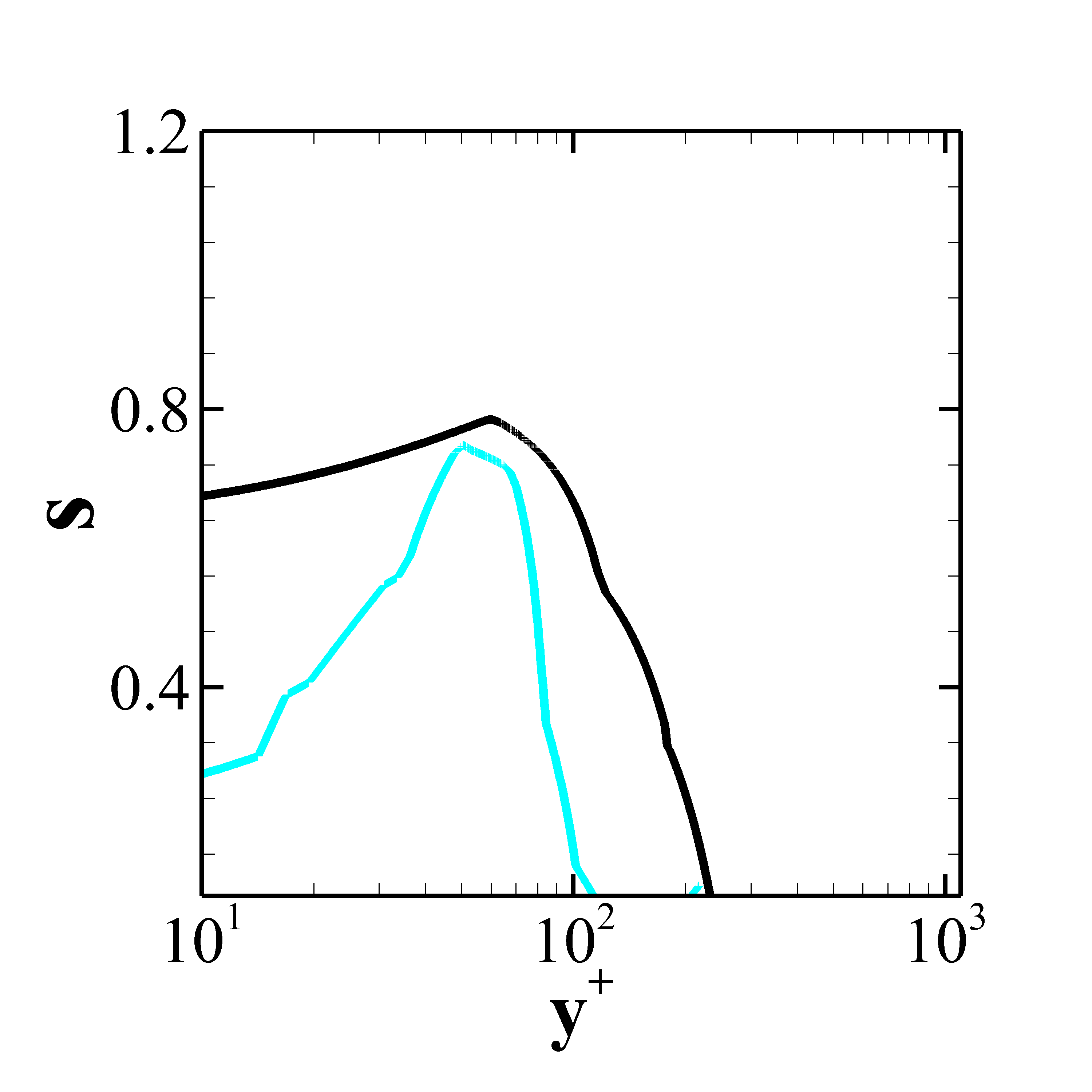}
\caption{}\label{fig:12b}
\end{subfigure}
\begin{subfigure}[b]{0.3\linewidth}
  \includegraphics[width=4cm,height=5cm,trim=50 50 50 50,clip]{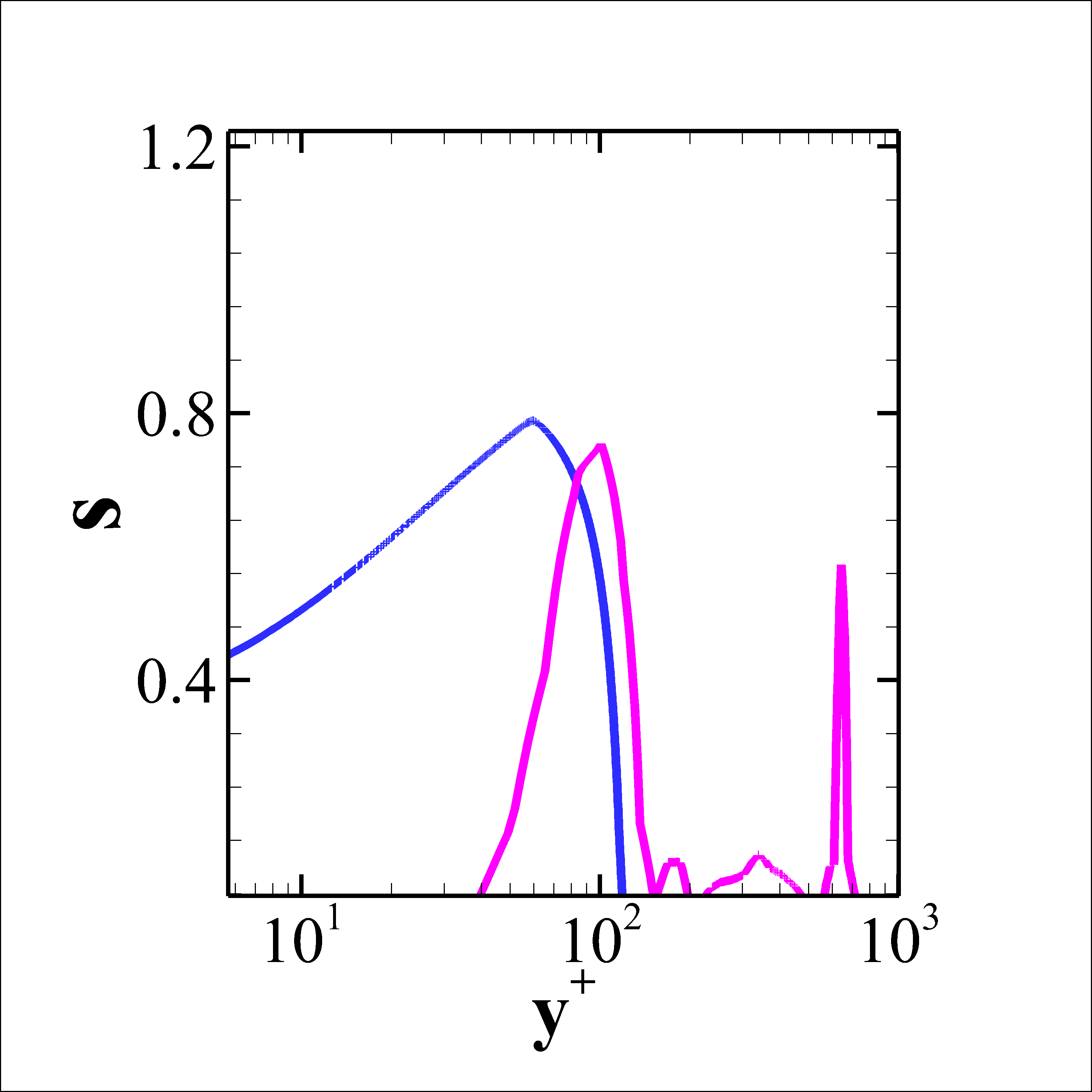}
  \caption{}\label{fig:12c}
\end{subfigure}
\caption{SGS activity parameter 's' for BFS-I}\label{fig:12}
\end{figure}

\quad Figure \ref{fig:12} reports the SGS activity parameter for all the models considered herein. The value predicted by the Dynamic Smagorinsky and the WALE model in the vicinity of the wall is higher as compared to the value predicted by the Dynamic WALE model. This is due to the higher value of eddy viscosity predicted by both the models. The molecular dissipation is higher near the wall and decreases as we move away from the wall whereas the dissipation due to the turbulent fluctuations follows the reverse trend. This combined effect results in a peak value of s in the inner boundary layer. As the outer boundary layer is reached, the dissipation due to the mean velocity field increases which results in a smaller peak. The peak value of s for the Dynamic Smagorinsky model in the Fig. \ref{fig:12a} and for the WALE model in the Fig. \ref{fig:12b} occurs much earlier than that for the Dynamic WALE model in the Fig. \ref{fig:12c}. Observation from the Fig \ref{fig:12b} confirms the non-existence of the second peak in the WALE model predictions. The small SGS dissipation due to the mean velocity field predicted by the WALE model seems to be responsible for this. 

\begin{figure}[H]
\begin{subfigure}[b]{0.3\linewidth}
\includegraphics[width=4cm,height=5cm,trim=50 50 50 50,clip]{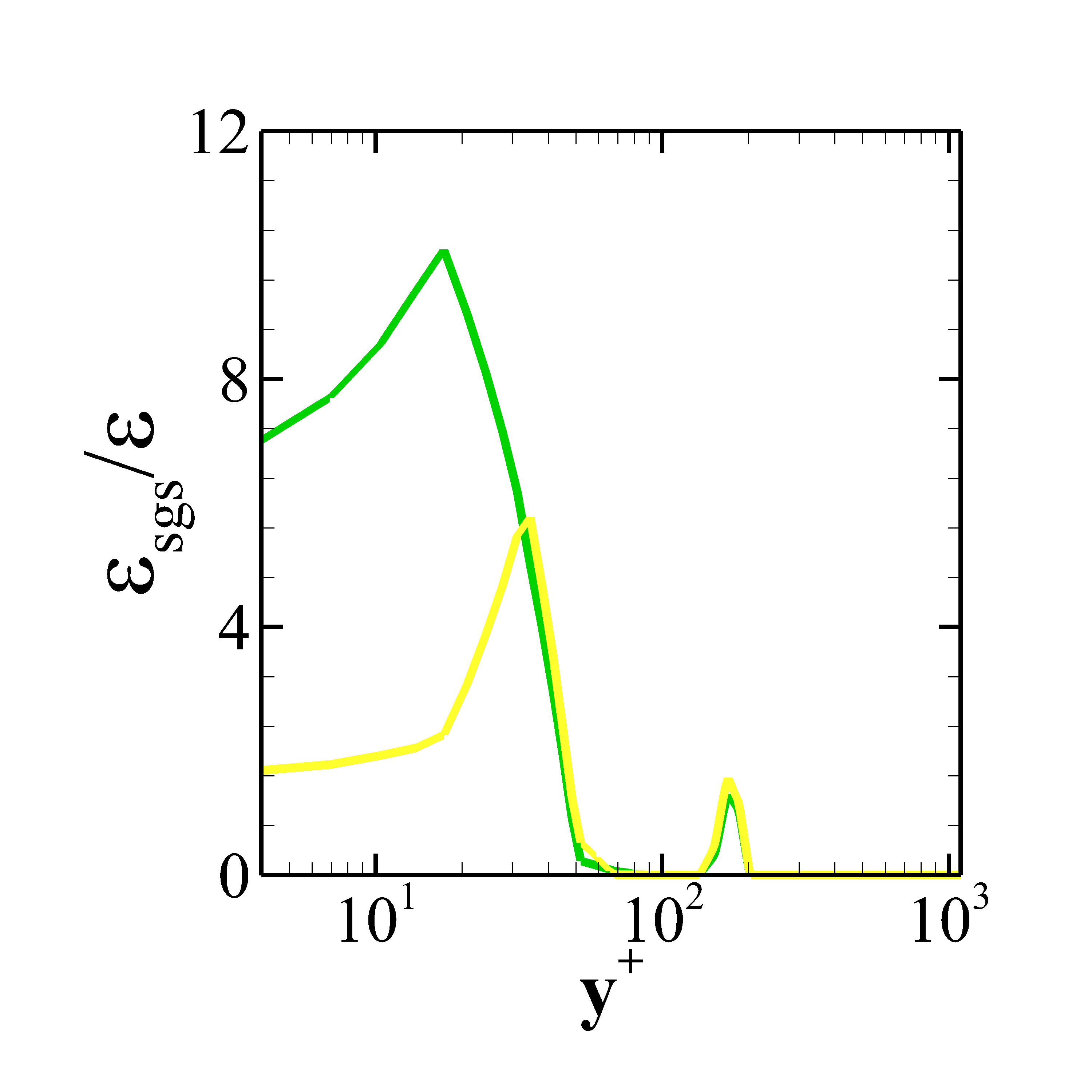}
\caption{}\label{fig:13a}
\end{subfigure}
\begin{subfigure}[b]{0.3\linewidth}
\includegraphics[width=4cm,height=5cm,trim=50 50 50 50 ,clip]{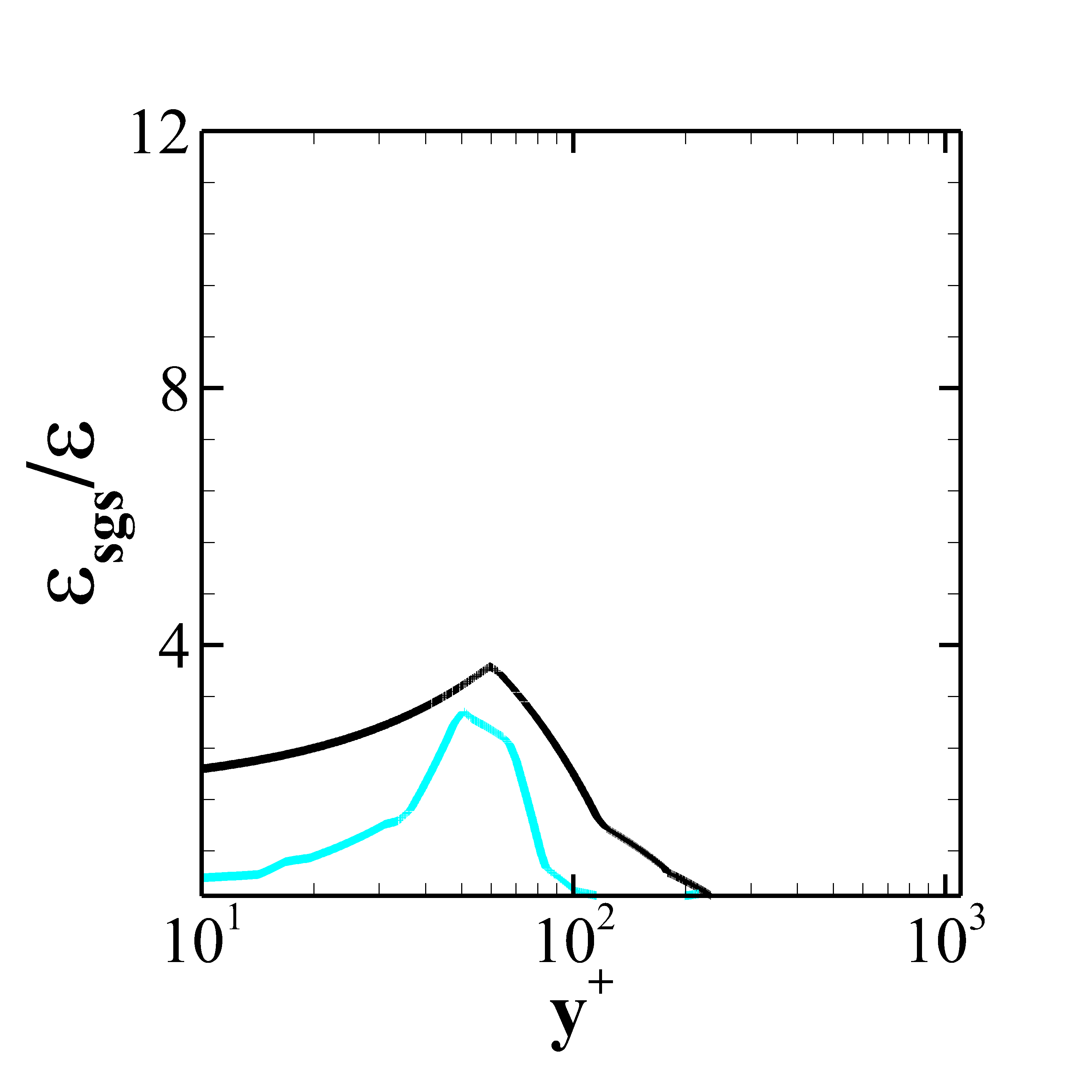}
\caption{}\label{fig:13b}
\end{subfigure}
\begin{subfigure}[b]{0.3\linewidth}
  \includegraphics[width=4cm,height=5cm,trim=50 50 50 50,clip]{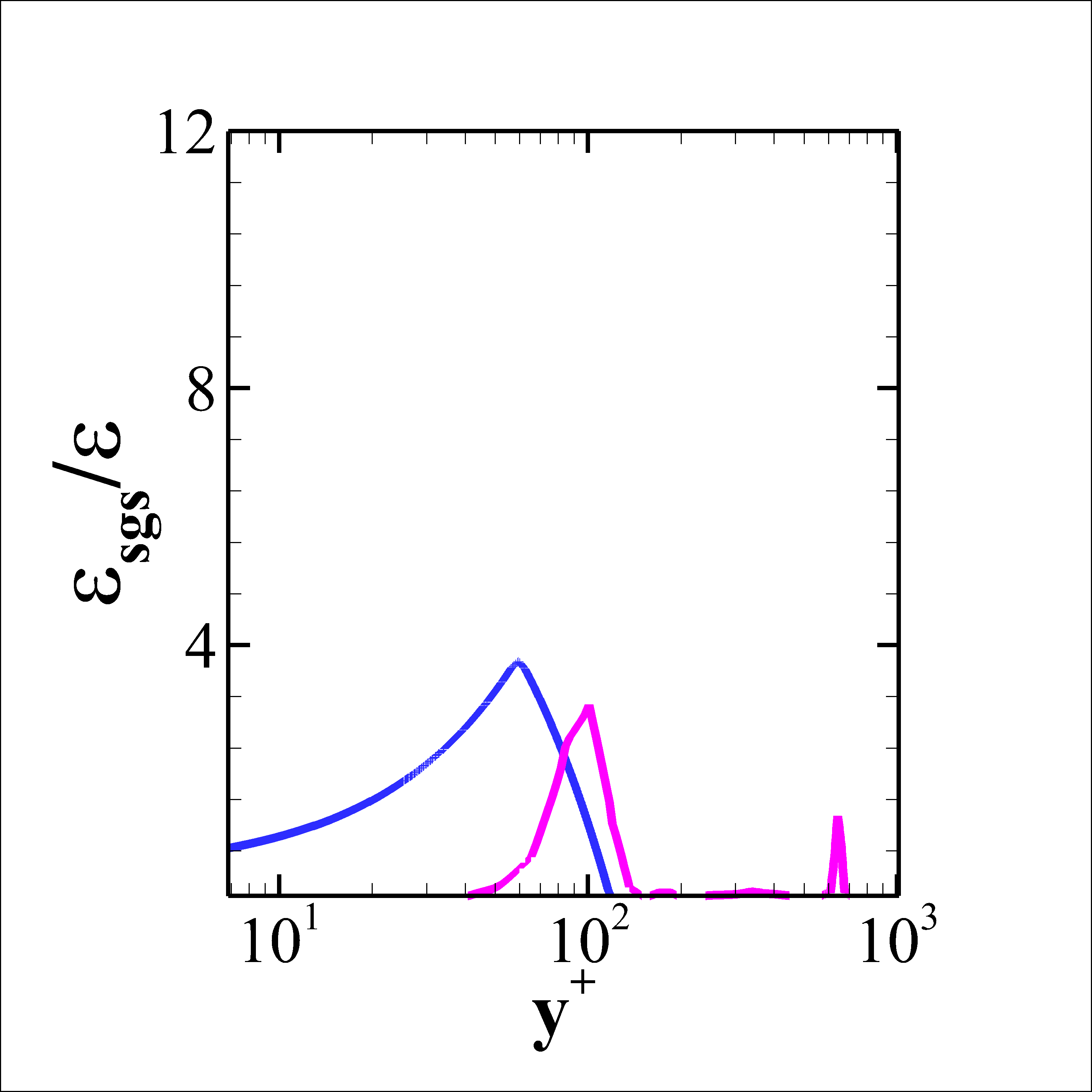}
  \caption{}\label{fig:13c}
\end{subfigure}
\caption{Ratio of SGS to Molecular Dissipation for BFS-I}\label{fig:13}
\end{figure}

\quad Now, Fig. \ref{fig:13} depicts the ratio of the SGS dissipation to the molecular dissipation. The fine grid for all the models predicts a lower value of the total SGS dissipation. The highest dissipation is provided by the coarse grid Dynamic Smagorinsky model. Further, Fig. \ref{fig:13c} shows that there is a contribution of the SGS dissipation due to the mean velocity field at the outer boundary layer for the Dynamic WALE model. For the WALE model this second peak is negligible which means that the dissipation due to the mean field contributes insignificantly to the total SGS dissipation. Also, the SGS dissipation for the Dynamic WALE model is very less as compared to the molecular dissipation in the near wall region. Thus, one can infer that the Dynamic WALE model provides acceptable trends for the total SGS dissipation.

\quad The supersonic BFS case is discussed in the next section.

\subsection{BFS-II}
\begin{figure}[H]
\centering
\includegraphics[width=12cm,height=5cm, trim=10 10 10 10, clip]{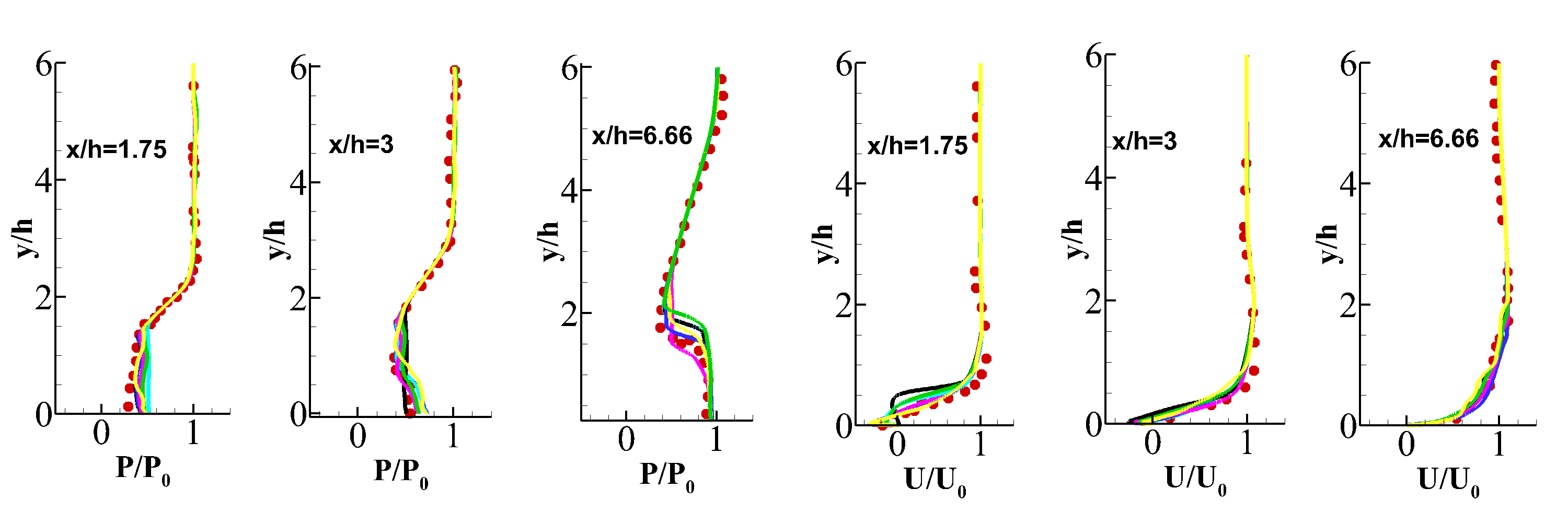}
\caption{Mean Streamwise Velocity and Pressure for BFS-II}\label{fig:14}
\end{figure}

\quad The supersonic flow over a BFS involves the flow separation at the step corner generating an expansion fan followed by reattachment at the downstream point which gives rise to a reattachment shock \cite{donaldson1967separation, hama1968experimental, takahashi2000self}. Thus, this flow geometry offers a way to test various models for a flow undergoing sudden acceleration and deceleration. The mean streamwise velocity and the pressure profiles at different axial locations are presented in the Fig. \ref{fig:14}.The profiles are validated against the experimental observations of McDaniel et al. \cite{mcdaniel1991staged}. The recirculation lengths predicted by various models are presented in Table  \ref{tab:4} which does not include the actual value of the reattachment length as it was not available in the experiment. The results show that the reattachment lengths predicted by the Dynamic WALE model are close to each other. This is also reflected in the profiles of pressure and velocity predicted by the Dynamic WALE model which closely follow each other. The pressure inside the recirculation region  is more or less constant which is evident from the pressure profiles at $x/h=1.75$ and at $x/h=3$ . There is an over prediction of pressure and velocity values  by the Dynamic Smagorinsky and the WALE model. At all the axial locations, the values predicted by the Dynamic WALE model closely follow the experimental trends. As the experiment did not report any turbulent stresses, we do not report them here too. However, we report the SGS activity parameter study for the present case in the next section.

\subsubsection{SGS activity parameter}
\begin{figure}[H]
\begin{subfigure}[b]{0.3\linewidth}
\includegraphics[width=4cm,height=5cm]{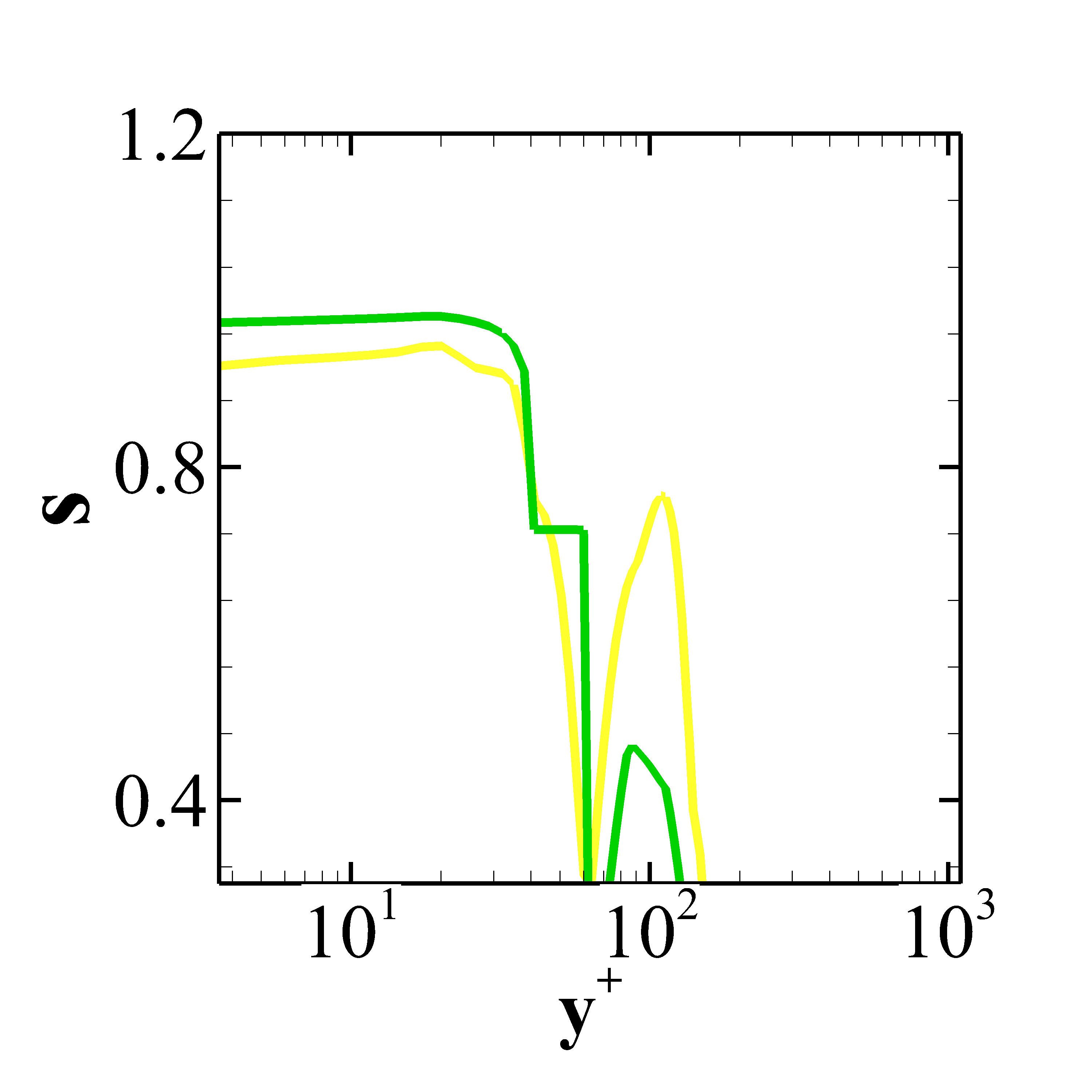}
\caption{}\label{fig:15a}
\end{subfigure}
\begin{subfigure}[b]{0.3\linewidth}
\includegraphics[width=4cm,height=5cm]{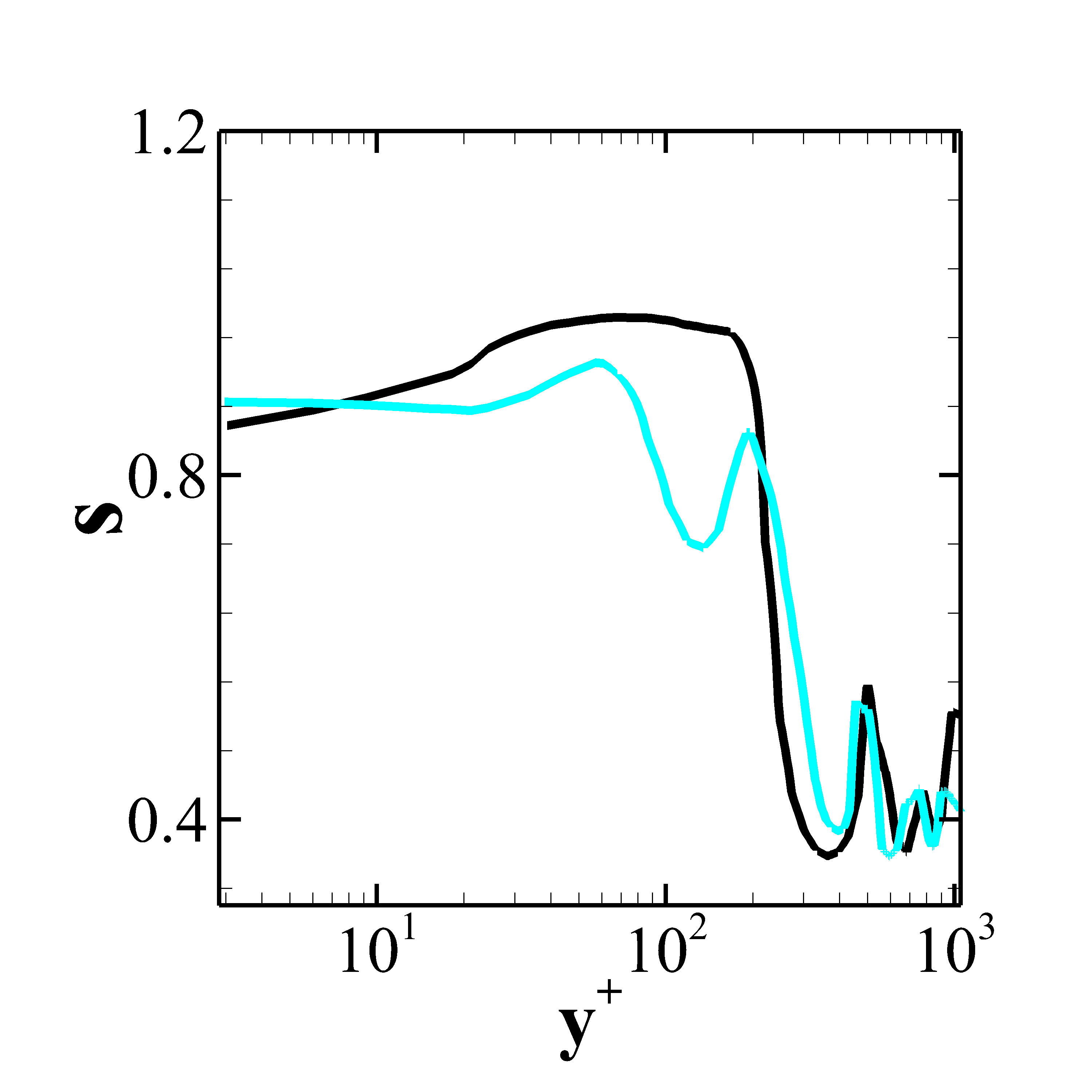}
\caption{}\label{fig:15b}
\end{subfigure}
\begin{subfigure}[b]{0.3\linewidth}
  \includegraphics[width=4cm,height=5cm]{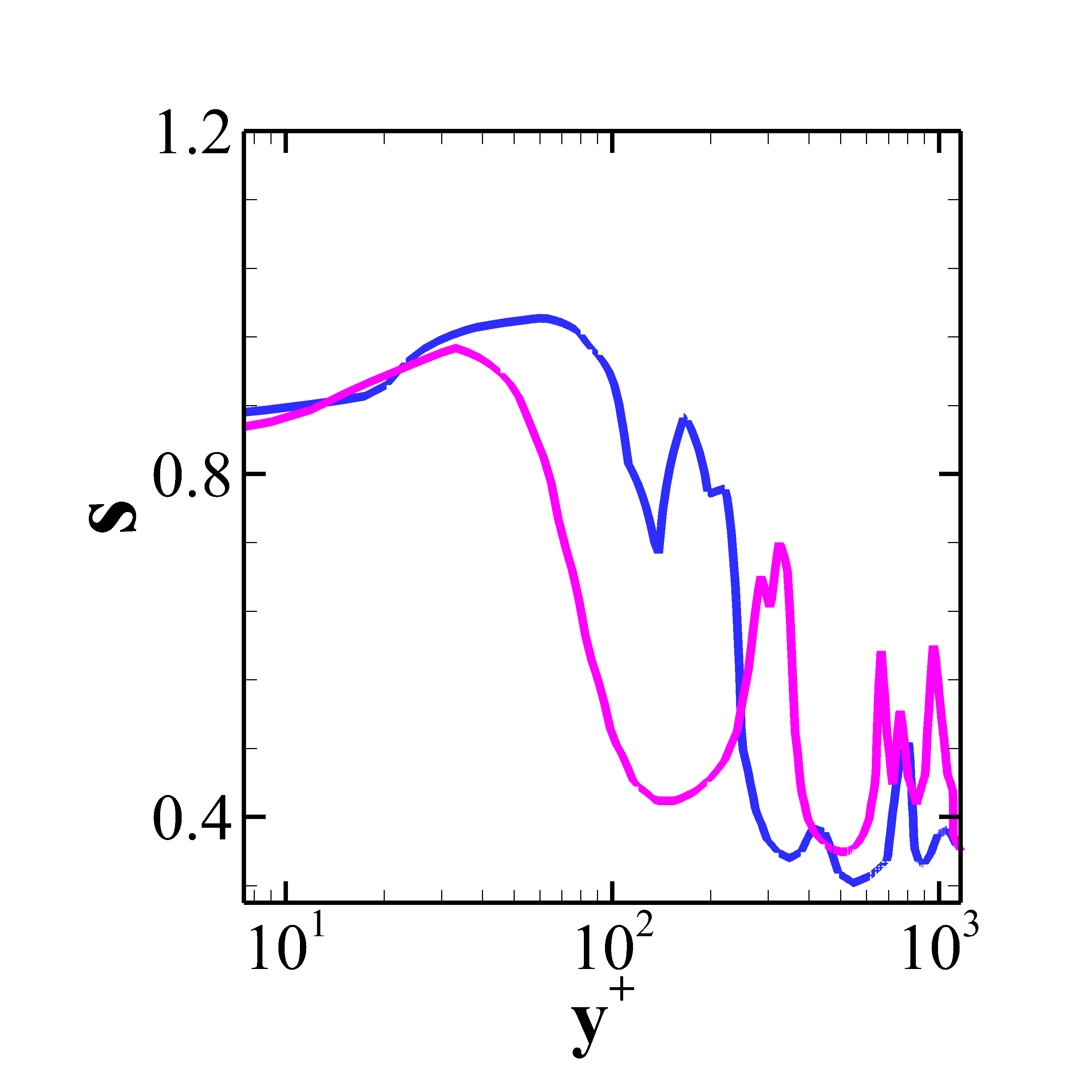}
  \caption{}\label{fig:15c}
\end{subfigure}
\caption{SGS activity parameter 's' for BFS-II}\label{fig:15}
\end{figure}

\begin{figure}[H]
\begin{subfigure}[b]{0.3\linewidth}
\includegraphics[width=4cm,height=5cm,trim=50 50 50 50,clip]{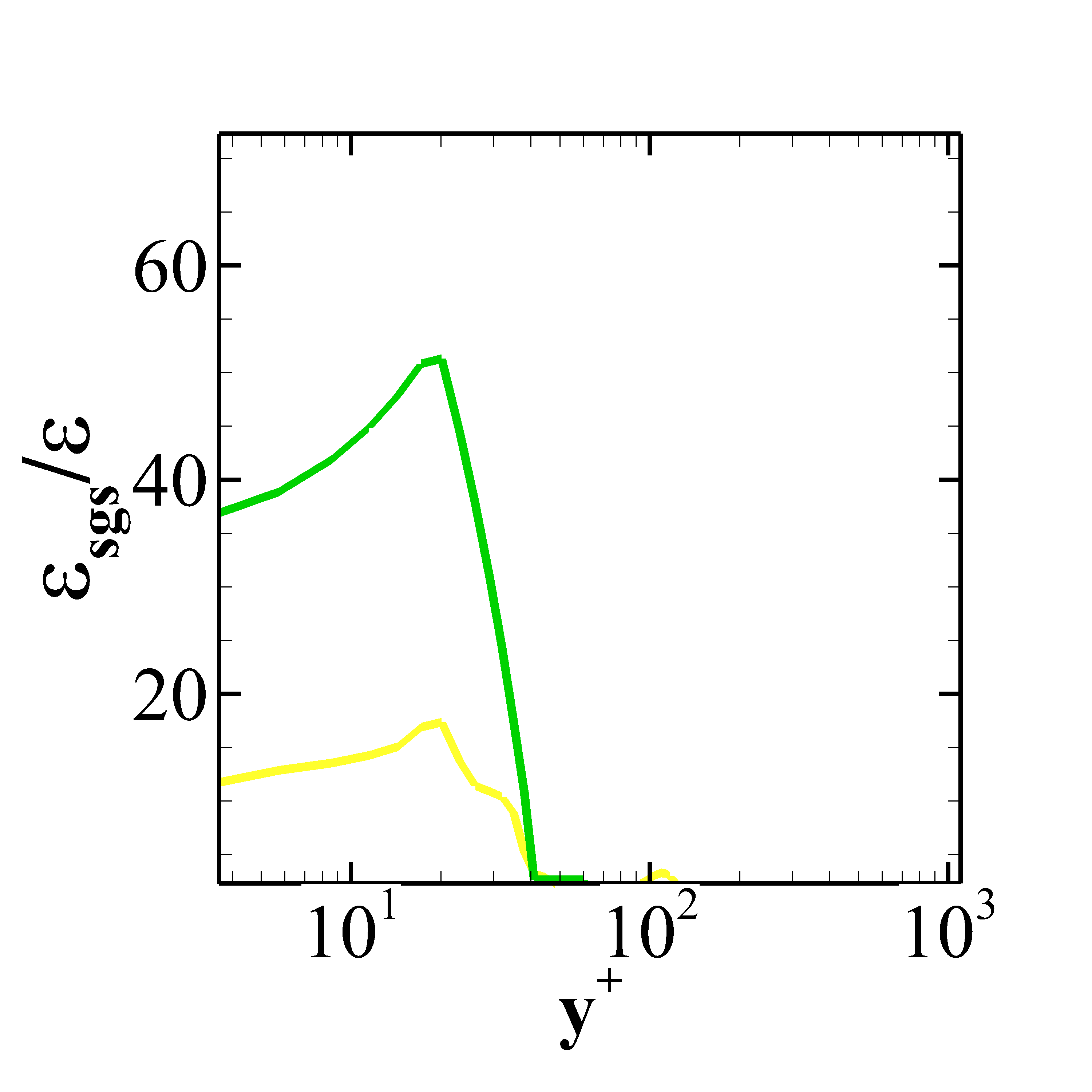}
\caption{}
\end{subfigure}
\begin{subfigure}[b]{0.3\linewidth}
\includegraphics[width=4cm,height=5cm,trim=50 50 50 50 ,clip]{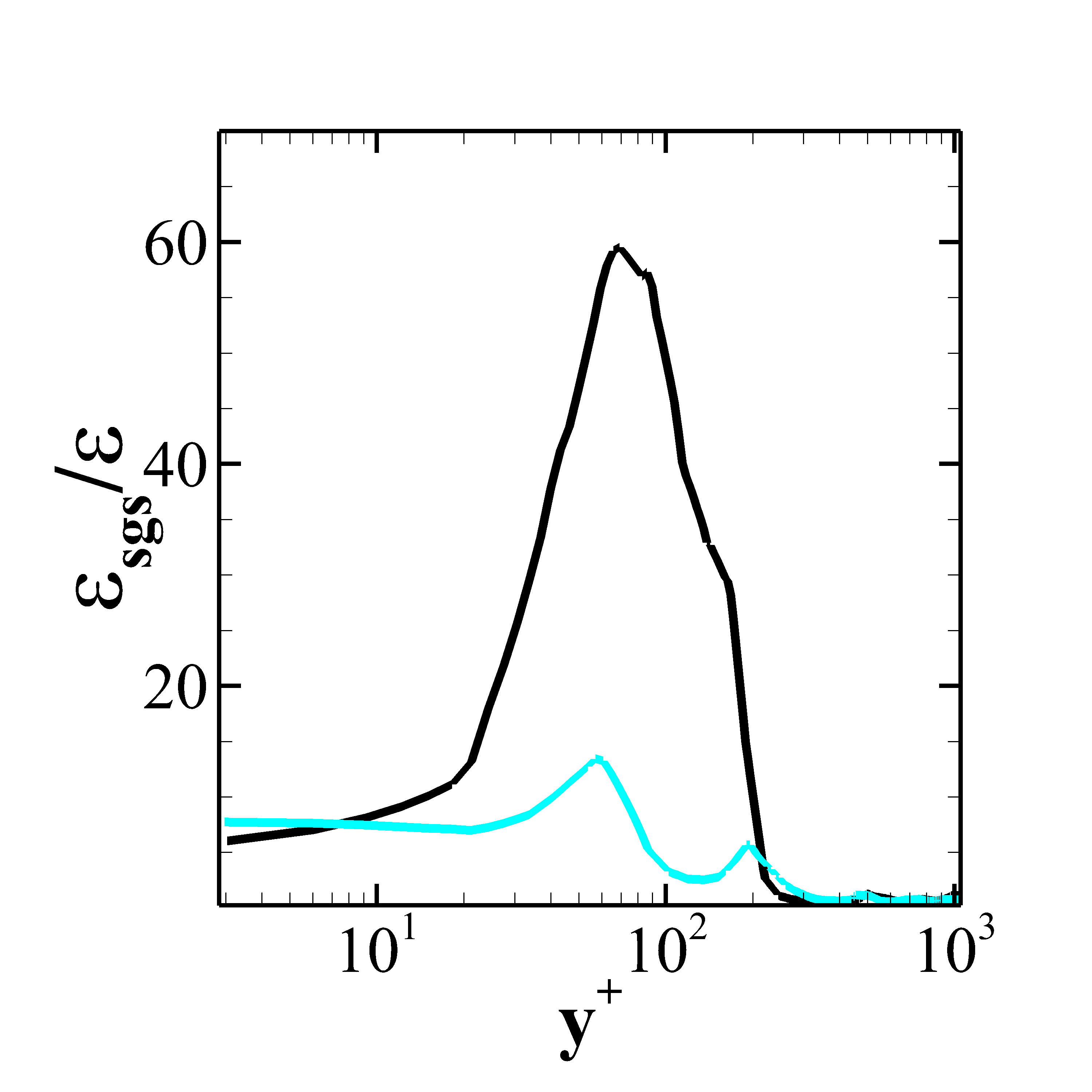}
\caption{}
\end{subfigure}
\begin{subfigure}[b]{0.3\linewidth}
  \includegraphics[width=4cm,height=5cm,trim=50 50 50 50,clip]{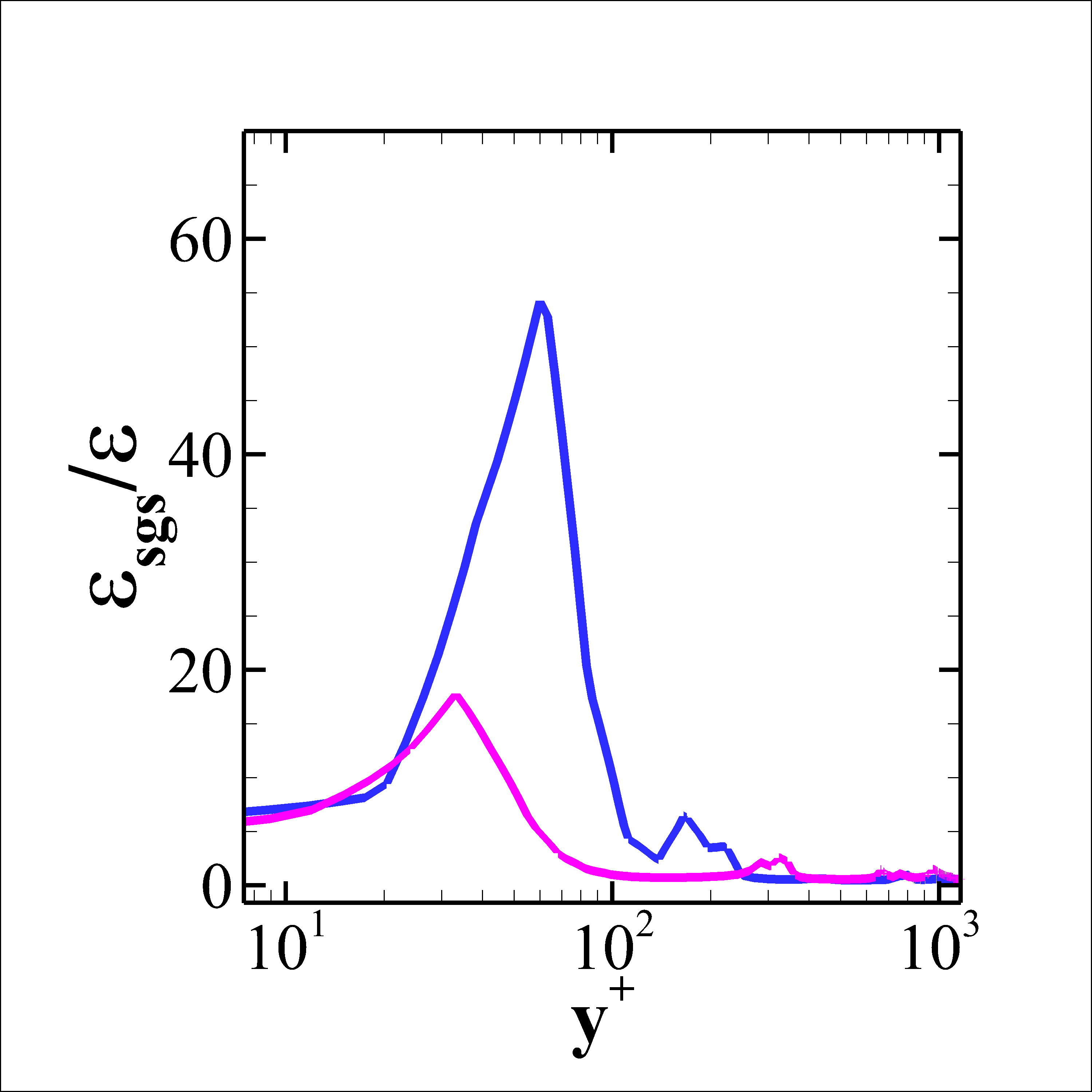}
  \caption{}
\end{subfigure}
\caption{Ratio of SGS to Molecular Dissipation for BFS-II}\label{fig:16}
\end{figure}

\quad Unlike the BFS-I case, the computational domain for the present case is not big enough due to the high Reynolds number which poses restrictions on the cost of the simulation for maintaining the desired $y^+$. Therefore, the  profiles for the calculation of SGS activity parameter are obtained at $x/h=14$. Though the flow downstream of the reattachment in this case as well behaves like a flow over a flat plate but the adverse pressure gradient experienced by the flow has more pronounced effects at this location. Figure \ref{fig:15} reports the SGS activity parameter for the different models where, Fig. \ref{fig:15a} reveals that the second peak, as discussed in the section for BFS-I, occurs near $y^+=100$ whereas for both WALE and Dynamic WALE it occurs near the outer region of the boundary layer which is evident from the Fig.\ref{fig:15b} and Fig. \ref{fig:15c}. However, for both the models, multiple peaks are observed after the first primary peak. This feature can be a result of many factors. The molecular viscosity changes across the boundary layer due to the change in temperature. Also, the contribution of the dissipation by turbulent fluctuations and the dissipation by mean velocity field are different for each model. On top of these factors, the effect of the adverse pressure gradient and the reattachment shock may also play a crucial role which is not trivial to quantify. The dissipation ratio profiles presented in the Fig. \ref{fig:16} also display multiple peaks for both the WALE and the Dynamic WALE model. Unlike the BFS-I case, the difference between the coarse grid and fine grid predictions is high. In the present case both the WALE and the Dynamic WALE model provide acceptable trends. 
 
\begin{table}[H]
\centering
\caption{Reattachment lengths($x/h$) for different models for BFS-II}\label{tab:4}
\begin{tabularx}{\linewidth}{| X | X |}
\hline
Dynamic Smagorinsky (coarse) & 4.75 \\ 
\hline
Dynamic Smagorinsky(fine) & 4.40 \\
\hline
WALE (coarse) & 5.20\\
\hline
WALE (fine) & 4.40 \\
\hline
Dynamic WALE (coarse) & 4.50 \\ 
\hline
Dynamic WALE (fine) & 4.60 \\ 
\hline
\end{tabularx}
\end{table}

\quad The results for the confined swirling flow case are presented in the next section. 
\subsection{CSF}
\begin{figure}[H]
\centering
\includegraphics[width=12cm,height=8cm, trim=10 10 10 10, clip]{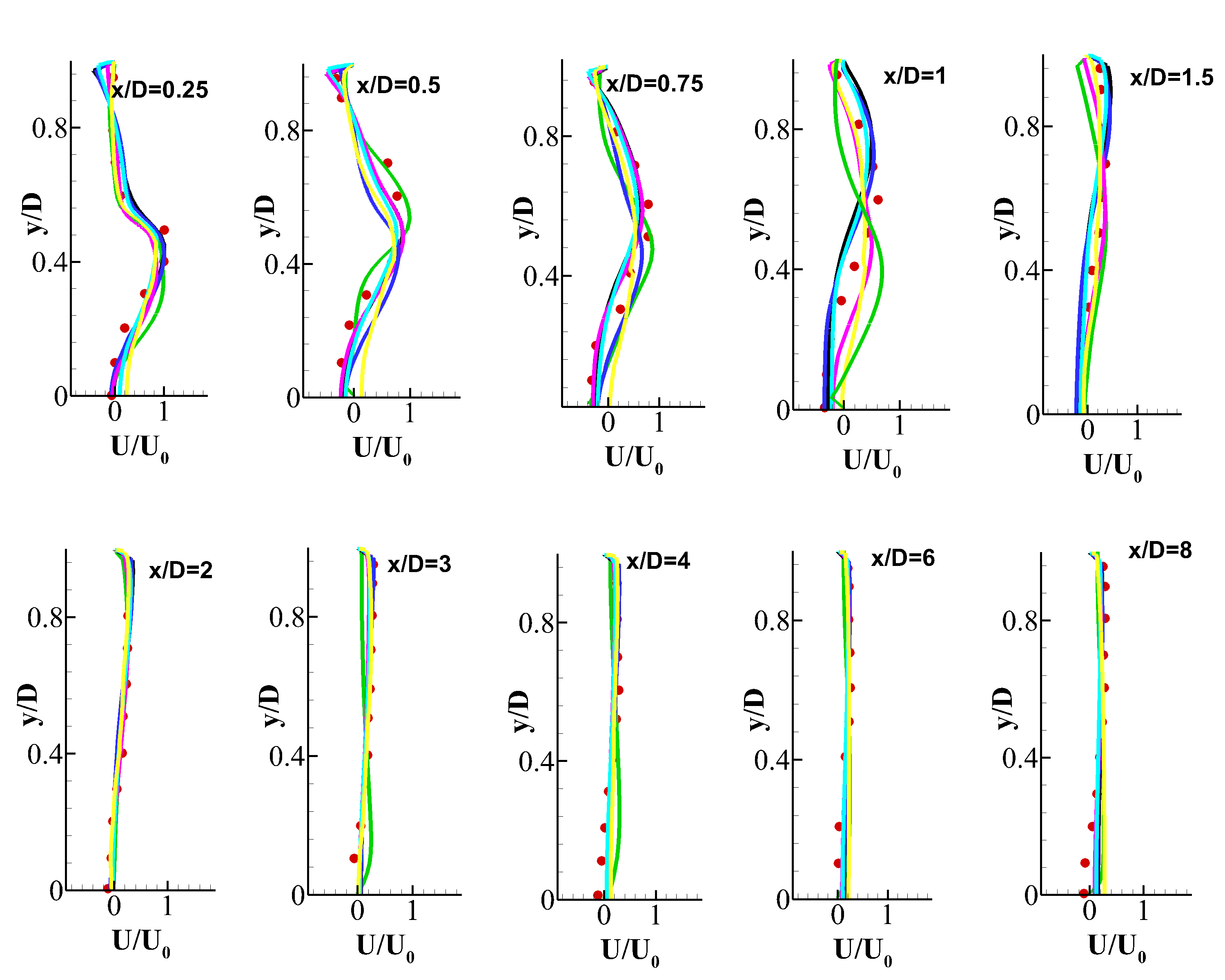}
\caption{Mean Streamwise Velocity profiles for CSF}\label{fig:17}
\end{figure}

\quad Similar to the BFS flow, a confined swirling flow experiences a flow separation when it suddenly encounters a tube with a wider diameter. Consequently, in the separation region, the central fluid moves radially outwards due to the expansion and the decrease in the axial velocity \cite{lopez1990axisymmetric, wang2004large, dellenback1988measurements}. This outward movement of the fluid results in the generation of an Internal Recirculation Zone (IRZ). It is a highly three dimensional recirculating region formed due to the fluid motion away from the influence of the wall. The mean velocity and turbulence stresses are validated against the experimental findings of Dellenback et al. \cite{dellenback1988measurements}. The mean streamwise velocity at different axial locations is presented in the Fig. \ref{fig:17}. It can be observed that the Dynamic Smagorinsky model fails to predict the separation well which is evident from the profiles at $x/D=0.25$ and $x/D=0.5$. The profiles of Dynamic Smagorinsky model are again far from the experimental trends at $x/D=0.75$ and $x/D=1$. This is due to the incorrect prediction of the IRZ. The Dynamic Smagorinsky model under predicts the wall normal extent of the IRZ.while the WALE model and the coarse grid Dynamic WALE model over predict its stretch. The Dynamic WALE model predicts separation well. However, it slightly underpredicts the wall normal stretch of the IRZ. The velocity profile at $x/D=1.5$ shows that the axial length of the IRZ is slightly larger for fine grid Dynamic WALE model and the Dynamic Smagorinsky model while it is smaller for other cases.

\begin{figure}[H]
\centering
\includegraphics[width=12cm,height=8cm, trim=10 10 10 10, clip]{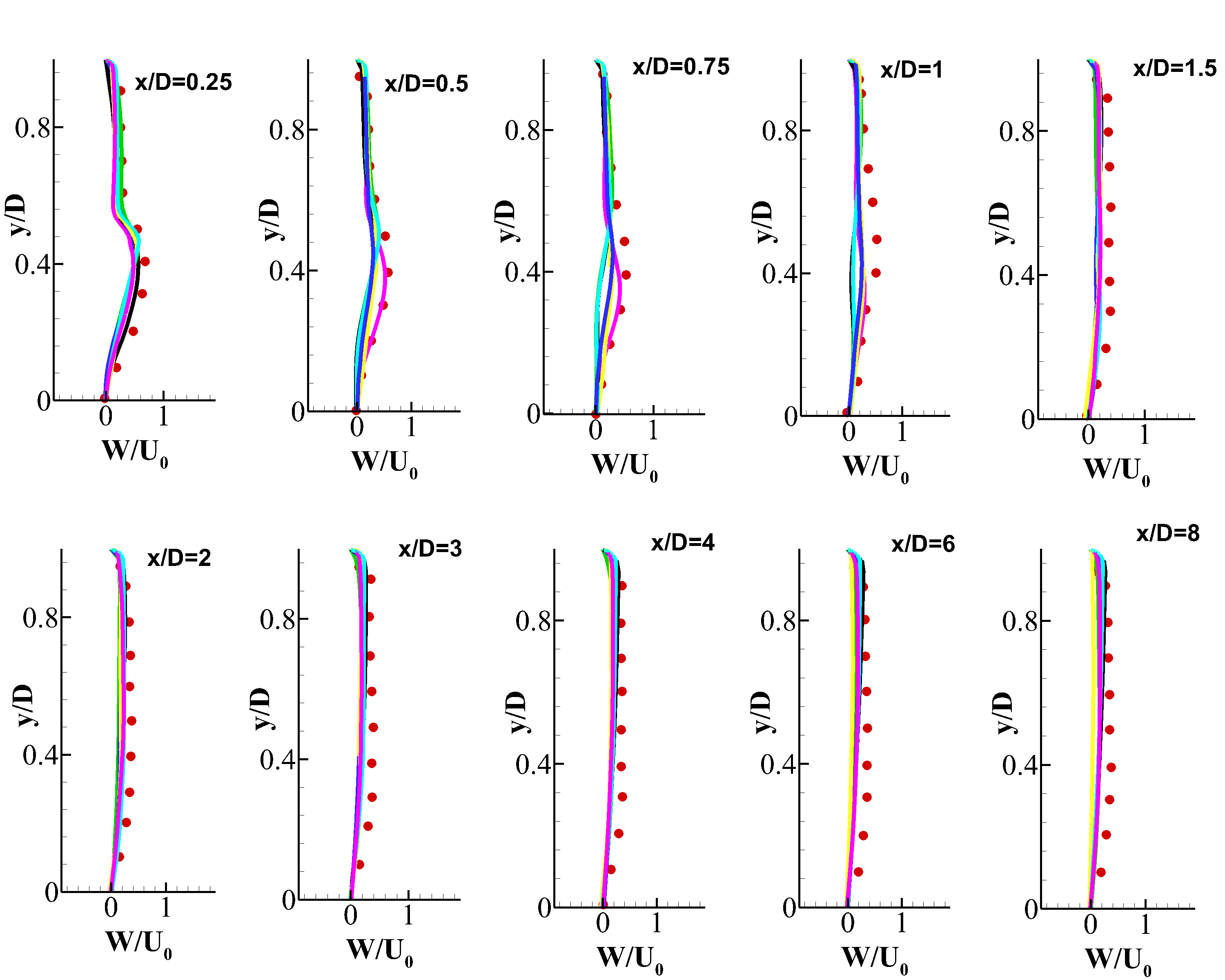}
\caption{Mean azimuthal Velocity profiles for CSF}\label{fig:18}
\end{figure}

\quad For completeness of the understanding of IRZ, Fig. \ref{fig:18} reports the mean azimuthal velocity component at different axial locations. The values predicted by the Dynamic WALE model are close to the experimental trends at the beginning of the IRZ ($x/D=0.5$ and $x/D=0.75$ . However, the value at the core of the IRZ is under predicted by all the models. This means that all the models underpredict the the extent of the IRZ in the spanwise direction.

\begin{figure}[H]
\centering
\includegraphics[width=12cm,height=8cm, trim=10 10 10 10, clip]{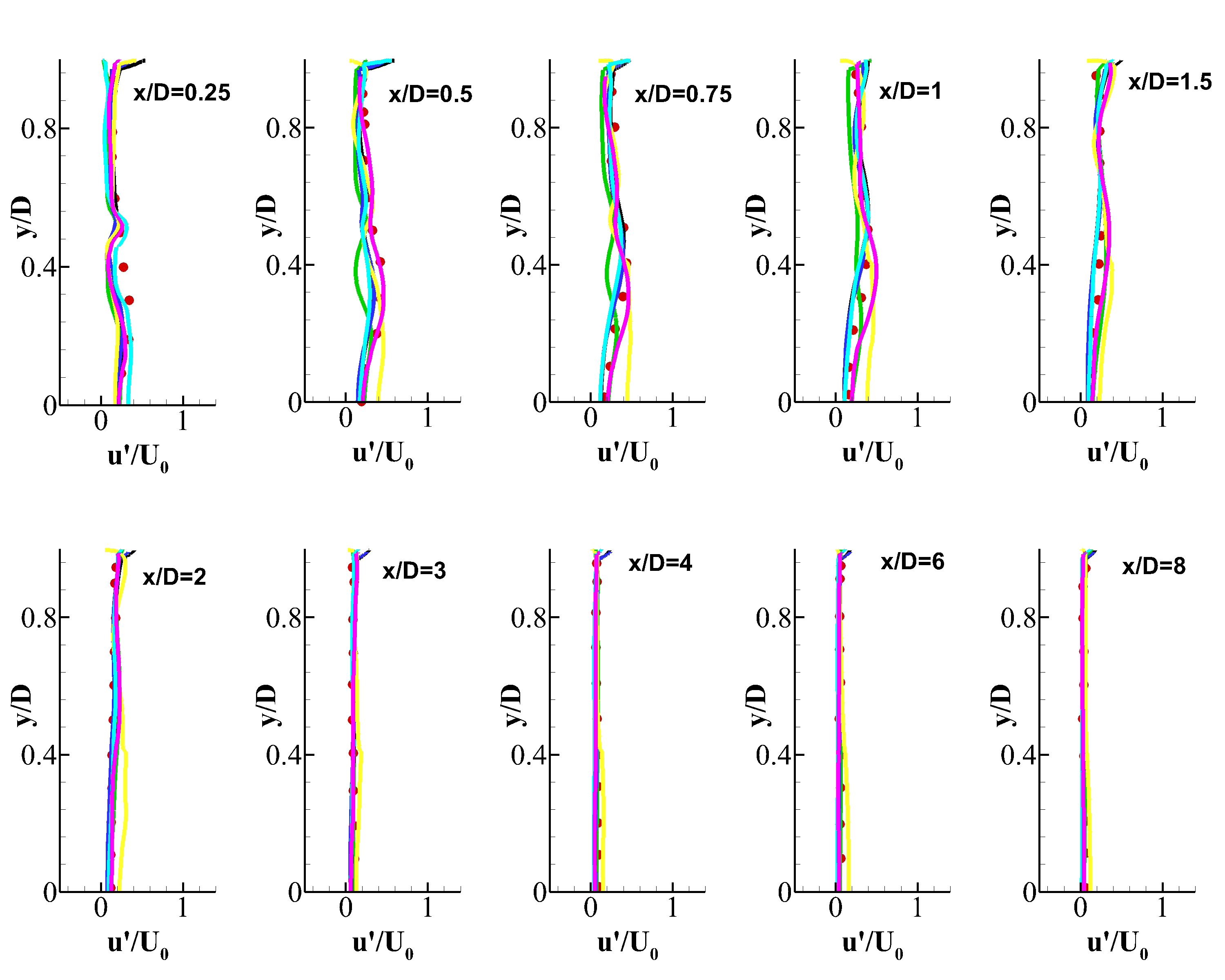}
\caption{Axial component of SGS Stress profiles for CSF}\label{fig:19}
\end{figure}

\quad Moreover, Fig. \ref{fig:19} presents the axial component of the SGS stress. The profiles at $x/D=0.25$ and $x/D=0.5$ display slight turbulence production at two locations. These correspond to the two shear layers formed due to the interaction of the flow walls of the swirler and with the IRZ. The fact that the WALE and the Dynamic WALE models show reasonable agreement with the experimental observations at these locations corroborates the earlier observation that the separation as well the starting of the IRZ are predicted well by these models. Both these models also show good agreement of the axial stress with the experimental findings at other axial locations as well.

\begin{figure}[H]
\centering
\includegraphics[width=12cm,height=8cm, trim=10 10 10 10, clip]{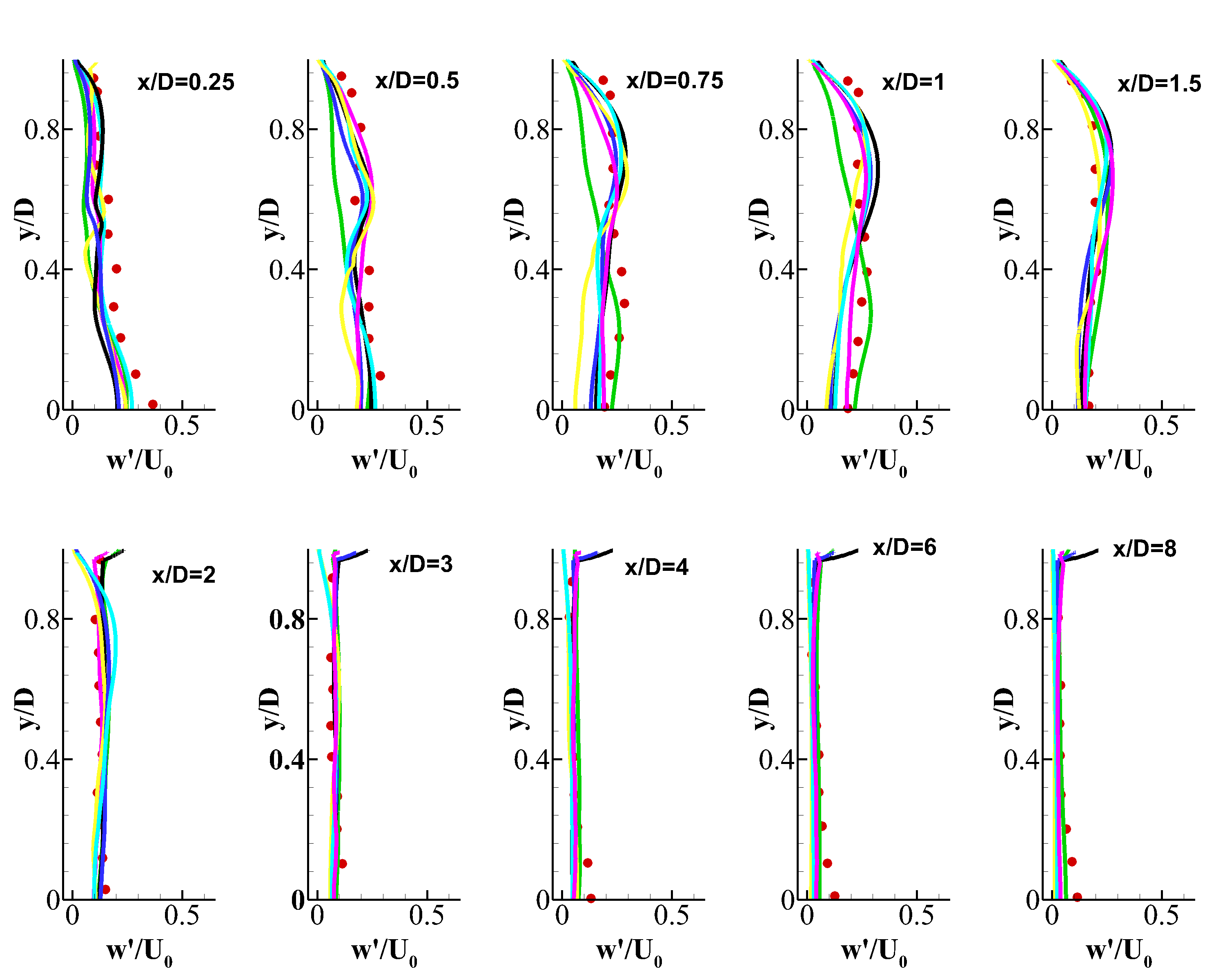}
\caption{Azimuthal component of SGS Stress profiles for CSF}\label{fig:20}
\end{figure}

\quad Though the mean azimuthal velocity predicted by all the models at $x/D=0.25$ are close to the experimental values, the azimuthal stress (Fig. \ref{fig:20}) is under predicted by all the models at this location.At other locations, the azimuthal stress profiles for the WALE and the Dynamic WALE model are closer to the experimental findings, though they are not in a perfect agreement. It again highlights the fact that the spanwise extent of the IRZ is not predicted well.

\quad Thus, the predictions of the mean velocities as well as the turbulent stresses are reasonably well for the WALE and Dynamic WALE models. However, both the models face difficulty in capturing the highly three dimensional IRZ formed at the center. Theoretically, the WALE or the Dynamic WALE model consider the rotation as well as the deformation rate while calculating the eddy viscosity. However, observation in the present study yields that for any flow involving a

  high level of three dimensionality, the turbulent shear stress is underpredicted. One way to overcome this limitation is to model the turbulent kinetic energy and the turbulent shear stress separately.  

\quad We have seen that all the geometries involved in the present study are associated with recirculation regions. The correct prediction of separated regions is important for an accurate description of the overall flowfield. Therefore, a comparison of separated regions for all the cases is done in the next section.
  
\subsection{Recirculation regions (Separated regions)}
\begin{figure}[H]
\centering
\includegraphics[width=12cm,height=8cm, trim=10 10 10 10, clip]{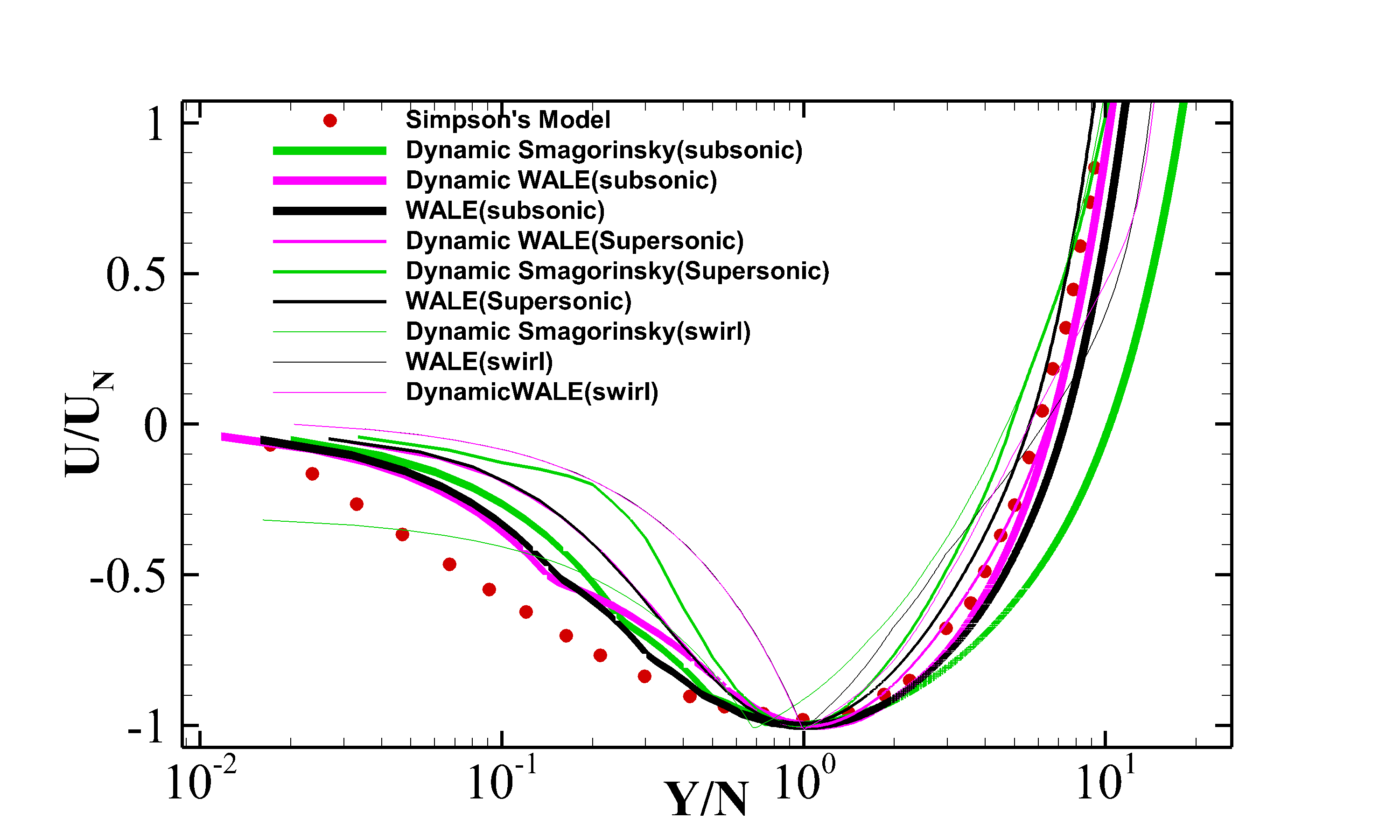}
\caption{Mean Streamwise Velocities in the recirculation regions}\label{fig:21}
\end{figure}

\quad The attached boundary layer for different flow configurations can be compared with a universal law of the wall. However, for a separated region, all the concepts of the boundary layer, by definition do not hold. Simpson et al. \cite{simpson1983model}  proposed an empirical relation for separated regions in a boundary layer with an adverse pressure gradient given by the following relation:
\begin{equation}
\frac{U}{|U_N|} = A \left[ \frac{y}{N} - log\left(\frac{y}{N}\right) - 1 \right] - 1
\end{equation}
where N is the distance from the wall to the point where maximum negative velocity occurs which is given by $U_N$. The value of the constant A is 0.3 as used in the Simpson et al. data. The profiles from the recirculation regions formed near the solid walls for all the geometries are collapsed on the empirical data. The profile is extracted at the center of the recirculation region for each case. The prediction of the WALE and the Dynamic WALE models for the BFS-I case are closest to the empirical results as observed from the Fig. \ref{fig:21}. The results for the BFS-II case, though better than the CSF case are still far from the empirical data. This is probably due to the acceleration of the flow by the expansion fan at the step corner which distorts the recirculation region for the BFS-II case. The profiles for the CSF case are not in agreement with the empirical observations. This is because the confined swirling flows are associated with highly three dimensional nature of the recirculation regions inside the domain. The conclusion is that the WALE and the Dynamic WALE perform better for separated flows in BFS case, although they have limitations when the separated regions have three dimensional nature.

\section{Conclusion}   
\quad The main aim of the present work is to study the effect of the grid sensitivity on the performance of the wall adapting SGS models for practical engineering flows over relatively coarser grids. This is done to check whether these models could be used to study the flow over complex geometries where resolving the complete wall would be a humongous task regarding the computational cost. It is observed that the profiles of mean velocity and turbulent stress predicted by the  Dynamic WALE model are in good agreement with the experimental observations for subsonic flow over the backward facing step. Though the results provided by the WALE model are not in good agreement with the experimental trends, the near wall profile of the eddy viscosity is reasonably close to to the $O(y^3)$, contrary to the eddy viscosity profile by the Dynamic WALE model which is slightly away from the desired trend. It is argued that this may be attributed to the Shear and Vortex Sensor or the cut off value used for the application of the dynamic procedure. It is also observed that the Dynamic WALE model provides best predictions for the SGS dissipation due to the turbulent fluctuations as well as the SGS dissipation due to the mean velocity field. The mean pressure and the mean streamwise velocity profiles for the supersonic flow over the backward facing step are also best predicted by the Dynamic WALE model. The SGS dissipation predicted by WALE and the Dynamic WALE model for this case follows acceptable trends. The Dynamic WALE model predictions are in close agreement with the experimental results for the axial velocity and axial SGS stress profiles for the confined swirling flow case. However, the mean azimuthal velocity and the azimuthal SGS stress are incorrectly predicted. It has been established that the construction of the velocity scale in the WALE or the Dynamic WALE model enables them to take care of the rotation as well the deformation rate in the flow. However, for the present case involving high swirl, the difference between the azimuthal and the axial components of the SGS stress is high which suggests that both the components must be modeled separately. The separation regions for all the cases are studied using the Simpson's empirical relation. The best results are predicted by the WALE and the Dynamic WALE model for the subsonic flow over the backward-facing step. The difference between the coarse and the fine grid predictions of the Dynamic Smagorinsky and the WALE model is high for all the cases. Since the quality parameters suggest that the grids are sufficiently resolved for a proper LES, this difference is due to the intrinsic nature of these models which inhibit them to accurately represent the flow physics on coarser grids. On the other hand, the Dynamic WALE model is almost insensitive to the grid resolution. Thus, it is concluded that the Dynamic WALE model behaves nicely for different geometries as well as for different flow regimes. It is a promising model and with some modifications, it is highly suitable to study the flow around complex geometries.

\section*{Acknowledgement}
The authors would like to thank the IIT Kanpur computer center \url{(http://www.iitk.ac.in/cc)} for providing the resources for performing the computational work.

\section*{References}

\bibliographystyle{elsarticle-num}
\bibliography{mybibfile}

\end{document}